\documentclass[twocolumn]{IEEEtran}

\usepackage{amsmath}
\usepackage{amsthm}
\usepackage{amsfonts}
\usepackage{amssymb}
\usepackage{cases}
\usepackage{cite}
\usepackage{graphicx}
\usepackage{mathrsfs}
\usepackage{subfigure}
\usepackage{epstopdf}
\usepackage{color}
\usepackage{caption}

\usepackage{enumitem}
\usepackage{setspace}
\usepackage{bm}
\newtheorem{prop}{Proposition}

\allowdisplaybreaks[4]

\begin{document}

\title{Joint Spatial Division and Diversity for Massive MIMO Systems
\author{Ke-Wen Huang, ~Hui-Ming Wang, ~\IEEEmembership{Senior Member, IEEE}, \hspace{0.05in}
	Jia Hou, \hspace{0.05in}  and Shi Jin,  ~\IEEEmembership{Senior Member, IEEE}}
}
\maketitle

\begin{abstract}
We propose a downlink beamforming scheme that combines spatial division and orthogonal space-time block coding (OSTBC) in multi-user massive MIMO systems. The beamformer is divided into two parts: a pre-beamforming matrix to separate the users into different beams with no interference between each other, which is designed based on the low rank covariance matrix of the downlink channel, and a linear precoding matrix using partial or even no channel state information (CSI) concatenated by an OSTBC. To construct the pre-beamforming matrix, a simple method that selects columns from DFT matrix is presented. To design the linear precoding matrix
with partial CSI of the effective channel after the pre-beamforming, we solve an
optimization problem to minimize the pairwise error probability (PEP) of the users under an individual power or sum power constraint, respectively. For the individual power constraint, a semi-definite relaxing (SDR) method with a sufficient condition achieving the globally optimal solution  is proposed to provide a performance benchmark. In addition, an efficient iterative successive convex approximation (SCA) method is provided to achieve a suboptimal solution. Furthermore, closed form solutions are derived under some special cases.
For the sum power constraint, we consider two different designs, i.e., minimizing the average PEP and minimizing the maximum PEP of all users. We find that both non-convex problems have a similar structure, and proposed a unified SCA-Alternating Direction Method of Multipliers (ADMM) algorithm to handle them. The SCA-ADMM method can be implemented in a parallel manner, and thus is with great efficiency. Simulation results show the
efficiency of our proposed JSDD scheme and the optimization method.

\end{abstract}
\begin{IEEEkeywords}
Massive MIMO, spatial division, orthogonal space-time block code, partial channel state information, SDR method, ADMM method.
\end{IEEEkeywords}

\section{Introduction}
Massive multiple-input and multiple-output (MIMO) is widely recognized as a critical technique in the future 5G wireless communications systems \cite{ScaleUPMIMO,MassiveMIMONG,Noncoop,EnergySpec}. With a large number of antennas, the channel vectors of multiple users are nearly orthogonal, and thus the uncorrelated multi-user interference and thermal noise go to zeros by simple transmit or receive techniques, e.g. MRT or MRC \cite{ScaleUPMIMO}, \cite{Noncoop}, which greatly increases the energy and spectral efficiencies  \cite{EnergySpec}. In addition, massive MIMO is also a promising technique to secure the wireless transmissions in the physical layer \cite{SecureMassive,ANDAS2,ANDAS}.

\subsection{CSIT for Massive MIMO downlink}
The great potential of massive MIMO dependents heavily on the perfect channel state information (CSI) at the transmitter  (generally BS), i.e., CSIT. Since the number of antennas is very large, the CSI acquisition brings  a heavy burden. To reduce the overhead, channel reciprocity is used in a time division duplex (TDD) system to obtain the downlink channel information from the uplink training sequence. However, channel reciprocity is always impaired by many imperfect factors, which leads to CSI errors and thus extra calibration is required to suppress the imperfections of channel reciprocity. On the other hand, frequency division duplex (FDD) massive MIMO system is an even more severe challenge, where the channel reciprocity does not hold anymore.
Furthermore, significant performance
degradation exists in practice due to channel estimation error,
feedback delay, and quantization error.

To address this challenge, one afford is to exploit the \emph{low rank property} of the downlink channel. In a massive MIMO system, (i) the antenna spacing of the antenna array is usually as small as half wave-length in order to keep the whole array aperture small; (ii) BS with large-scale antenna array has to be elevated at the top
of high buildings such that there are few local scattering. Due to these reasons, the channel spatial spread of angles of departure (AOD) at the BS is always narrow \cite{covMatSumForm,covMatForm}, which leads to a spatially correlated downlink channel. Mathematically, the independently and identically distributed (i.i.d.) channel assumption violates and the rank of the channel covariance matrix is significantly lower than the number of transmit antennas. Therefore, by exploiting this low-rank property, tremendous studies have been focused on the topic of downlink channel estimation \cite{covMatSumForm,massiveMIMOchannelEst,MassiveMIMOchannelGao}, simplified CSI feedback schemes \cite{Massivemimolimitedfeedback,W.Shen2017ICC,W.Shen2018TCOM} and downlink signal and transmission scheme designs \cite{covMatForm,massiveMIMOFDD,BVMC}. In this paper, we concentrate on the downlink transmission scheme designs.

\subsection{Joint spatial division and multiplexing (JSDM) scheme}
In \cite{covMatForm}, a  joint
spatial division and multiplexing (JSDM) scheme for the FDD system is proposed. The downlink beamforming is divided into two stages: a pre-beamforming matrix based on the channel second-order statistics, and a multi-user MIMO (MU-MIMO) precoding matrix based on the efferctive channel coefficeints. Under the JSDM scheme, the users are divided into  $K$ groups, and each user group consists of $J$ users.
Due to the low-rank property, after pre-beamforming, the effective channel realizations of the users in each user group have significantly reduced dimensions compared with the number of antennas. Then the subsequent MU-MIMO precoding is designed based on the downlink training and the CSI feedback of the instantaneous effective channel realization of each user group with $J$ users, the overhead of which could be reduced by a factor of 10.
In a word, \emph{spatial division (pre-beamforming) is exploited to separate users into groups and spatial multiplexing (MU-MIMO precoding) is utilized to send data streams to multiple users in each group}. A similar idea has been adopted for users with multiple antennas in \cite{BVMC}.

Nevertheless, after pre-beamforming, the downlink training and the CSI feedback of the instantaneous effective channel realizations of multiple users
are still required. When the effective channel information is not perfectly estimated and fed back to the BS, the performance of the system will be greatly impaired, due to the sensitivity of the beamforming technique to the CSIT, such as in the JSDM scheme proposed in \cite{covMatForm}.

On the other hand, we know that
space-time coding (STC) technique is designed to provide transmit diversity gain when the transmitter with multiple antennas does not have the CSIT \cite{Alamouti,SpaceTimeCodes,SpaceTimeOrthgonal}.
Among various kinds of STCs,
orthogonal space-time block codes (OSTBC) received widely concerns due to its full diversity property with a linear decoding complexity. \footnote{
It is well known that Maximum likelihood (ML) decoding to achieve the full diversity gain is equivalent to linear decoding for OSTBCs. Furthermore, some advanced designs have also provided full diversity gain with linear receivers instead of ML decoding, such as zero-forcing and MMSE receivers \cite{STClinearReceiver1}, \cite{STClinearReceiver}. They get rid of the  orthogonal structure to improve the OSTBC rate further.}  In addition, if some certain but not perfect CSIT is available, partial channel information could be utilized to improve the OSTBC performance further \cite{STBCAndBeamforming,DesignChannel}. It has been shown that this strategy outperforms the conventional beamforming (without perfect CSIT) as well as conventional OSTBC significantly.

\subsection{Our proposed scheme}
The low rank property of the spatial channel  and the robustness of the OSTBCs to accurate CSIT motivate us to combine them together to release the burden of CSIT acquisition in massive MIMO systems, and to provide sufficient spatial multiplexing and diversity gain for multiple users.
In this paper, we propose a new transmission scheme that combines the spatial division with the OSTBC in the massive MIMO downlink transmissions. In our scheme, for each user group, the BS schedules only one user at each time, and therefore, the BS simultaneously serves $K$ users. The basic idea is to permit $K$ users to access the downlink transmissions via spatial division by utilizing the low-rank property of channels, and for each user OSTBC with partial CSI (estimated CSI with errors) is utilized to provide diversity gain.  Therefore, we name our scheme as \emph{joint spatial division and diversity} (JSDD) scheme.

Specially, the beamformer in the proposed JSDD transmission scheme is composed of two parts: (i) a pre-beamforming matrix to spatially separate the users and eliminate the inter-user interference, and (ii) a linear precoding matrix combining the OSTBCs of the users to achieve diversity gain.
For the pre-beamformer, we design a DFT matrix based eigen-beamforming scheme.
More specifically, for the uniform linear array in the massive MIMO transmitter, the covariance matrix of the channel can be well approximated by a circular matrix as the number transmit antenna gets large \cite{covMatForm,massiveMIMOFDD,BVMC}. Hence, the eigenvectors of the channel covariance matrix forms a DFT matrix. Due to the low rank property, we propose a low complexity method to directly find those columns from the DFT matrix to form the eigenvectors matrix. The eigenvector matrix is the corresponding pre-beamforming matrix when multiple users have non-overlapped angular spreads. After pre-beamforming, the users are spatially separated and for each user the dimension of the effective channel has been greatly reduced, which significantly releases the channel acquisition overhead burden. In the second part of the proposed JSDD scheme, a linear precoding matrix for transmitting the OSTBCs of the users is designed and optimized by utilizing the partial CSIs of the effective channels, where we take the imperfection of effective channel acquisition into consideration.
The optimization objective is to minimize the pairwise error probability (PEP). In such a way, the JSDD scheme has a light burden for the CSIT acquisition, and is very robust to the CSIT imperfection.

A critical problem in the JSDD scheme is how to optimize the linear precoding matrix, which is a non-convex problem. In the paper, we try to solve this problem under both the individual user power and sum power constraints, respectively. For the individual user power constraint, we first solve the problem using semidefinite relaxation (SDR) and Gaussian randomization method. In addition, a sufficient condition is presented to achieve the global optimum, i.e., the rank constraint is met. The SDR solution can be taken as a benchmark of the performance.
We then propose a successive convex approximation (SCA) based iterative optimization method to  obtain a suboptimal solution.
Furthermore, some special cases with closed-form optimal solutions are also discussed, including the hign SNR, low SNR and no effective CSIT cases. For the sum power constraint, two different performance metrics, i.e.,  min-max and average PEP of the system, are investigated.
An SCA method combined with ADMM algorithm is proposed to handle both of the two problems. The proposed method can be implemented in a parallel manner, and therefore is generally much more efficient than the SDR method.

\subsection{Application scenarios}
Compared with JSDM, the proposed JSDD scheme has reduced the number of data streams from $KJ$ to $K$, and thus the overall throughput is reduced. However, higher diversity gain could be achieved so the bit error rate is lower, which will be shown in the numeric part in this paper.
In many applications in 5G wireless communications, such as ultra reliable low latency communications (uRLLC), higher reliability is more important than larger data rate.
The JSDD scheme could be applied in various scenarios in massive MIMO downlink transmissions. For example, in a multiple-group multicast scenario. The BS is required to broadcast multiple data streams to multiple geographically separated  groups of users, respectively, where each group of users  co-located in a cluster are interested in a same file. Another example is  directional broadcasting a common message to a cluster of users in a same region with uRLL requirement.

The paper is organized as follows. In Section \ref{SecSystemModel}, the channel model is presented, where the channel representation and the covariance matrix approximated by the DFT matrix are proposed. The transmission scheme combining spatial division with OSTBC is proposed in Section \ref{TranScheme}. The individual user power constraint is addressed in Section \ref{indSection} while the sum power constraint in Section \ref{sumSection}. Simulation results are presented in Section \ref{simSection} and Section \ref{conclusionSection} concludes the paper.

\emph{Notation}: In this paper, we use the upper-case and lower-case boldface letters for matrices and vectors respectively. $(\pmb \cdot)^T$, $(\cdot)^H$, $(\cdot)^{-1}$, $\det(\cdot)$, $\mathrm{Re}(\cdot)$, $\mathrm{tr}(\cdot)$, $||\cdot||_F$, and $||\cdot||_2$ denote the transpose, conjugate transpose, inversion, determinant, real part, trace, Frobenius norm, and $l_2$ norm, respectively. $\pmb{I}_m$ denotes the $m\text{-by-}m$ identity matrix. $\mathbb{E}(\cdot)$ is the statistical expectation. $\mathrm{mod}(x,y)$ denotes modulus after $x$ divided by $y$. $\mathcal {CN}(\pmb{\mu}, \pmb{R})$ denotes the circularly symmetric complex Gaussian distribution with mean value and covariance matrix given by $\pmb{\mu}$ and $\pmb{R}$, respectively.

\section{Channel Model}\label{SecSystemModel}
\begin{figure*}[t]
\begin{center}
\includegraphics[height = 1.5 in]{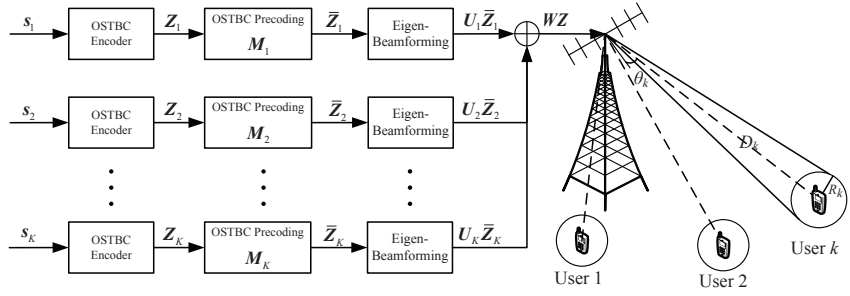}
\end{center}
\caption{System model.}\label{systemModel}
\rule{\textwidth}{.2mm}
\vspace{-7mm}
\end{figure*}

Fig. \ref{systemModel} depicts a massive MIMO downlink system, where a BS equipped with $M$ antennas serves $K$ single-antenna users simultaneously. In this paper, we consider the uniform linear array (ULA) with half wavelength interval between antennas at the BS, and in the user terminal, the one-ring scatter channel model is considered \cite{covMatSumForm,covMatForm,MicroMobile,FadingCor}, where the $k$-th user is surrounded by a ring of scatters of radius $R_k$ and the distance from the $k$-th user to the BS is $D_k$. The downlink channel vector $\pmb h_k\in\mathbb{C}^{M\times 1}$ of the $k$-th user can be expressed as \cite{covMatSumForm},
\begin{align}
\pmb h_k = \frac{1}{\sqrt Q}\sum_{q=1}^Q\alpha_{k,q}\pmb a(\theta_{k,q}),\quad k=1,2,\cdots,K,\label{oneRingChannel}
\end{align}
where $Q$ denotes the number of i.i.d. paths, and $\alpha_{k,q}\sim\mathcal{CN}(0,1)$ is the complex gain, which is independent over user index $k$ and path index $q$. Besides, $\pmb a(\theta_{k,q})$ is the steering vector with azimuth angle $\theta_{k,q}$ and has the form $
\pmb{a}(\theta_{k,q})= \left[1,e^{-j\pi\sin(\theta_{k,q}),\cdots,e^{-j\pi(M-1)\sin(\theta_{k,q})}}\right]^T.
$
BS equipped with a large number of antennas is elevated at a high altitude, say on the top of a high building, or a dedicated tower, such that there are few surrounding scatterers. In this case, the one-ring  model is a reasonable channel model, and the angular spread of the $k$-th user's channel is restricted within a narrow region $\left[\bar{\theta}_k-\Delta_k,\bar{\theta}_k+\Delta_k\right]$ with $\bar{\theta}_k\in[-90^o,90^o]$ the mean azimuth angle, and $\Delta_k\approx\arctan(R_k/D_k)$ the  angular spread of the $k$-th user's channel.
According to \cite{covMatForm}, the covariance matrix of the $k$-th user channel $\pmb R_{k}=\mathbb{E}(\pmb h_k\pmb h_k^H)$ can be calculated by the Toeplitz form
\begin{align}
[\pmb R_{k}]_{m,n} = \frac{1}{2\Delta_k}\int_{\bar{\theta}_k-\Delta_k}^{\bar{\theta}_k+\Delta_k}e^{-j\pi(m-n)\sin(\theta)}d\theta,\label{covarianceMat}
\end{align}
where $[\pmb R_{k}]_{m,n}$ denotes the entry in the $m$-th row, $n$-th column of $\pmb R_{k}$ with $m,n=1,2,\cdots,M$. We have $\pmb h_k\sim\mathcal{CN}(\pmb 0,\pmb R_{k})$ and the channel vector has a low rank expression as
\begin{align}
\pmb h_k=\pmb U_k\pmb \Lambda_k^{1/2}\pmb v_k,\label{KLRepresent}
\end{align}
where $\pmb v_k\in \mathbb{C}^{r_k\times 1}\sim\mathcal{CN}(\pmb 0,\pmb I_{r_k})$, $\pmb R_{k}=\pmb U_k\pmb\Lambda_k\pmb U_k^H$ is the eigenvalue decomposition of the covariance matrix,   $\pmb U_k\in \mathbb{C}^{M\times r_k}$ satisfies $\pmb U_k^H\pmb U_k=\pmb I_{r_k}$, $\pmb \Lambda_k=\mathrm{diag}\left(\lambda_{k,1},\lambda_{k,2},\cdots,\lambda_{k,r_k}\right)$ is a diagonal matrix with ordered eigenvalue as diagonal elements, i.e., $\lambda_{k,1}\geq\lambda_{k,2}\geq\cdots\geq\lambda_{k,r_k}$, and $r_k$ is the rank.

In a downlink massive MIMO  system, the acquisition of the channels $\{\pmb h_k\}_{k=1}^K$ at the BS is critical but also difficult due to the large $M$. However,  the covariance matrix $\pmb R_{k}$ possesses a low-rank property due to the high correlation among $\pmb a(\theta_{k,p})$'s, and the dimension of $\pmb v_k$ is much less than $M$, i.e., $r_k\ll M$  in \eqref{KLRepresent}.
Since the covariance matrix $\pmb R_{k}$ is determined by the azimuth angle and the angular spread, which is slowly changed and reciprocal for both FDD and TDD systems, an efficient way to estimate $\{\pmb h_k\}_{k=1}^K$ is to  calculate $\pmb R_{k}$ according to (\ref{covarianceMat}) first, and then do the eigenvalue decomposition to obtain $\pmb U_k$ and $\pmb\Lambda_k$, and finally estimate $\pmb v_k$ for several or tens of channel coherence time intervals.

However, the accompanied computational complexity involved in the eigenvalue decomposition of high-dimensional covariance matrix in a massive MIMO system is hardly affordable. { In \cite{covMatForm,covMatSumForm,PilotReuse,GaoffMassiveMIMO},  a DFT-matrix based method to calculate $\pmb U_k$ and $\pmb\Lambda_k$ is proposed,
namely, the columns of $\pmb U_k$ could be well approximated by some columns of $M$ dimension DFT matrix. To make the paper self-contained, we cite the following proposition to give a way to determine the indices of these columns.

\begin{prop}
\label{P1}
If the column index $i$ satisfies $I_{\min}^{(k)}\leq i\leq I_{\max}^{(k)}$, then the $i$-th column of DFT matrix should be selected as an approximate eigenvector of $\pmb{R}_{k}$, where $I_{\min,k}$ and $I_{\max,k}$  are
\begin{align}
\label{proposition}
\begin{aligned}
&I_{\min}^{(k)} = \mathop{\mathrm{argmin}}\limits_{i}
\left|\Xi_{\min}-\frac{2\pi(i-1)}{M}\right|, \\
&I_{\max}^{(k)} = \mathop{\mathrm{argmin}}\limits_{i}
\left|\Xi_{\max}-\frac{2\pi(i-1)}{M}\right|,
\end{aligned}
\end{align}
with $\Xi_{\min}\triangleq\mathrm{mod}\left(\pi\sin\left(\theta_{\mathrm{min}}^{(k)}\right),2\pi\right)$,
$\Xi_{\max}\triangleq\mathrm{mod}\left(\pi\sin\left(\theta_{\mathrm{max}}^{(k)}\right),2\pi\right)$,
$\theta_{k,\mathrm{min}}\triangleq\bar{\theta}_k-\Delta_k$ and $\theta_{k,\mathrm{max}}\triangleq\bar{\theta}_k+\Delta_k$.
Note that if $I_{\max,k}<I_{\min,k}$, for example $(\theta_{k,\mathrm{max}}>0)\cap(\theta_{k,\mathrm{min}}<0)$, then we should selected those columns with their column indices satisfying $(i\geq I_{\min,k})\cup(i\leq I_{\max,k})$.
\end{prop}
\begin{IEEEproof}
See the proofs in \cite{covMatForm,PilotReuse,GaoffMassiveMIMO}. The basic idea behind Proposition \ref{P1} is that the steering vectors of an ULA for different AoDs tend to be orthogonal when the number of antennas is large, see references such as \cite{W.Shen2017ICC,W.Shen2018TCOM}.
\end{IEEEproof}
}

Based on Proposition \ref{P1}, we simply select those columns of DFT matrix to form $\pmb U_k$, and multiply $\pmb R_{k}$ by these vectors as $\pmb U_k^H\pmb R_{k}\pmb U_k$ to get the eigenvalues.
In this paper, as depicted in Fig. \ref{systemModel}, we assume that the $K$ users who are simultaneously served by the BS are non-overlapped considering the angular spread. This can be guaranteed by letting the BS perform user scheduling according to the locations of the users. More specifically, following \cite{covMatForm}, we assume the BS divides multiple users into several user groups, which satisfies that the users in the same group have similar AoDs while for the users in different groups, their AoDs are significantly distinct. At each time, we assume that the BS selects one user from each user group to serve. Similar assumption has also been assumed in \cite{BVMC}.
Based on this assumption, when we use \eqref{proposition} to select columns from the $M$-dimension DFT matrix, the columns for different users are different, i.e., the eigenvectors of $K$ users are orthogonal to each other as
\begin{align}
\pmb U_k^H\pmb U_m = \pmb 0\quad k,m=1,2,\cdots,K,k\neq m. \label{character}
\end{align}
This property can be used to relieve the inter-user interference, which will be discussed in the following section.

\section{JSDD Scheme}\label{TranScheme}
Based on the channel model introduced in Section II, we now propose our beamforming with OSTBC transmission scheme for multi-user downlink communications. The basic idea is as follows:
First, a DFT-matrix based pre-beamforming is used to separate multiple users with nearly orthogonal eigenspace. After the pre-beamforming, the inter-user interferences are eliminated. For each user, the equivalent channel vector is now
 $\pmb v_k$, whose
dimension is greatly reduced from $M$ to $r_k$ according to (\ref{KLRepresent}). However, to obtain the accurate equivalent instant channel $\pmb v_k$, channel estimation and feedback are still required.
When the feedback channel information is not accurate, i.e., channel information is outdated or channel estimation  and feedhack have errors, the performance of the beamforming schemes will be greatly deteriorated, such as in the JSDM scheme proposed in \cite{covMatForm}. On the other hand, we know that OSTBC schemes do not require the CSI at the transmitter, which means that OSTBC is robust to CSI errors at the transmitter end. Motivated by this observation, we then propose to combine the conventional OSTBC with partial CSIT after the DFT-based pre-beamforming \cite{STBCAndBeamforming}. Specifically, the transmitted symbols are first encoded by OSTBC encoder and then a linear precoding matrix is exploited to minimize the  pairwise error probability (PEP) by using the partial equivalent downlink channel state information on $\pmb v_k$.
We now detail the JSDD scheme.

\subsection{JSDD scheme}
Let $s_{k,i}$, $k=1,2,\cdots,K$, $\ i=1,2,\cdots,L_k$, denote the $i$-th transmitted symbol to the $k$-th user, where $L_k$ is the number of symbols embedded in a OSTBC block for the $k$-th user. We assume that $\mathbb{E}|s_{k,i}|^2=1$ and $\mathbb{E}\left(s_{k,i}^R(s_{k,i}^I)^*\right)=0$ with $s_{k,i}^R$ and $s_{k,i}^I$ the real and imaginary part of $s_{k,i}$ respectively. Besides, we assume that the symbols for different users are independent. The OSTBC codeword matrix of the $k$-th user $\pmb Z_k\in \mathbb{C}^{N_k\times T_k}$ can be expressed as \cite{OSTBCForm},
\begin{align}
\pmb Z_k = \sum_{i=1}^{L_k}\pmb\Phi_{k,i}s_{k,i}^R+j\sum_{i=1}^{L_k}\pmb\Psi_{k,i}s_{k,i}^I,
\end{align}
where  $\pmb\Phi_{k,i},\pmb\Psi_{k,i}\in \mathbb{R}^{N_k\times T_k}$ are OSTBC precoding matrices with entries drawn from $\{-1,0,1\}$. For the OSTBC, the following two equivalent conditions holds true:
\begin{align}
1)&\quad \begin{aligned}
\pmb Z_k\pmb Z_k^H = \left(\sum_{i=1}^{L_k}|s_{k,i}|^2\right)\pmb{I}_{N_k},
\end{aligned}\\
2)&\quad \left\{\begin{aligned}
    \forall i: &\quad \pmb\Phi_{k,i}\pmb\Phi_{k,i}^H = \pmb{I}_{N_k},\pmb\Psi_{k,i}\pmb\Psi_{k,i}^H = \pmb{I}_{N_k}  \\
    \forall n\neq i:&\quad \left\{
    \begin{aligned}
    \pmb\Phi_{k,n}\pmb\Phi_{k,i}^H &= -\pmb\Phi_{k,i}\pmb\Phi_{k,n}^H,\\
    \pmb\Psi_{k,n}\pmb\Psi_{k,i}^H &= -\pmb\Psi_{k,i}\pmb\Psi_{k,n}^H,
    \end{aligned}\right.  \\
    \forall n,i: &\quad\pmb\Phi_{k,n}\pmb\Psi_{k,i}^H = \pmb\Psi_{k,n}\pmb\Phi_{k,i}^H.
\end{aligned}\right.\label{OSTBCForm}
\end{align}
In fact, the diversity gain that can be achieved by a certain kind of OSTBC scheme mainly depends on its order, i.e., $N_k$, and the number of channel usages that are required to transmit an OSTBC codeword matrix is given by $T_k$. The rate of such a OSTBC codeword is given by $L_k/T_k$. As shown in \cite{SpaceTimeOrthgonal}, \cite{SNRApp}, the rate-1 complex-symbol OSTBC exists only for $N_k=2$, which is the well-known Alamouti code. For $N_k=3, 4$, rate-3/4 OSTBC exists, while for $N_k>4$, only rate-1/2 has been constructed. Besides, the equivalent channel dimension is $r_k$, then we have $N_k\leq r_k$, which will be the rank constraint in the optimization problem. In this paper, to simplify the mathematical description, we assume that $T_1=T_2=\cdots=T_K=T$, however, the proposed method in this paper also suits to the case where $T_1,T_2,\cdots,T_K$ are different.


Stacking the OSTBC codeword matrices for all the users together yields $\pmb Z = \left[\pmb Z_1^T,\pmb Z_2^T,\cdots,\pmb Z_K^T\right]^T\in \mathbb{C}^{N\times T}$, where $N = \sum_{k=1}^KN_k$. The OSTBC codewords are then linearly weighted by a matrix $\pmb W\in\mathbb{C}^{M\times N}$ before the downlink transmission. We will derive the detailed form of $\pmb W$ in the following.
At the receiver, the received symbols of the $k$-th user are
\begin{align}
\pmb y_k = \pmb h_k^H\pmb {WZ}+\pmb e_k,\label{singleReceivedVector}
\end{align}
where $\pmb y_k\mathbb{C}^{1\times T}$ consists of the received symbols and $\pmb e_k\in \mathbb{C}^{1\times T}$ is the noise vector. The entries of $\pmb e_k$ are assumed to be i.i.d. complex Gaussian random variables with zero mean and variance $\sigma^2$.

Denoting $\pmb W=\left[\pmb W_1,\pmb W_2,\cdots,\pmb W_K\right]$ with $\pmb W_k\in \mathbb{C}^{M\times N_k},k = 1,2,\cdots,K$,  (\ref{singleReceivedVector}) can be rewritten as
\begin{align}
\pmb y_k = \pmb h_k^H\pmb W_k\pmb Z_k+\sum_{m=1,m\neq k}^K\pmb h_k^H\pmb W_m\pmb Z_m+\pmb e_k.\label{anotherSingle}
\end{align}
The first term in \eqref{anotherSingle} contains the intended symbols for the $k$-th user while the second term is the inter-user interference. Substituting \eqref{KLRepresent} into \eqref{anotherSingle}, for $k = 1,2,\cdots,K$, we have
\begin{align}
\pmb y_k & =\pmb v_k^H\pmb \Lambda_k^{1/2}\pmb U_k^H\pmb W_k\pmb Z_k \nonumber \\
&\quad +\sum_{m=1,m\neq k}^K\pmb v_k^H\pmb\Lambda_k^{1/2}\pmb U_k^H\pmb W_m\pmb Z_m+\pmb e_k.\label{noiseForm}
\end{align}
We can see that if we design $\pmb W_k$ to satisfy $\pmb U_k^H\pmb W_m = 0,\ m=1,2,\cdots,K,m\neq k$, then the inter-user interference vanishes. To satisfy this constraint, we design $\pmb W_k$ as
\begin{align}
\pmb W_k = \pmb U_k^\bot\pmb M_k,\label{subUnitaryRepre}
\end{align}
where $\pmb U_k^\bot\in\mathbb{C}^{M\times r_k}$ is the null space projection matrix of ${\pmb U}_{\bar k} \triangleq \left[\pmb U_1,\pmb U_2,\cdots,\pmb U_{k-1},\pmb U_{k+1},\cdots,\pmb U_K\right]$, i.e., ${\pmb U}_{\bar k}^H\pmb U_k^\bot=\pmb 0$, and $\left(\pmb U_k^\bot\right)^H\pmb U_k^\bot=\pmb I_{r_k}$, $\pmb M_k\in\mathbb{C}^{r_k\times N_k}$ is a matrix whose entries are our concern afterwards.

From (\ref{character}), if the eigenspace of the $K$ users has no overlap, which is equivalent to the case that the angle spread range of all the users are not overlap, the projection matrix can be chosen as $\pmb U_k^\bot = \pmb U_k$ directly. Then (\ref{noiseForm}) turns to be
\begin{align}
\pmb y_k
         &=\bar{\pmb v}_k^H\pmb M_k\pmb Z_k+\pmb e_k,\label{partialCSIChannel}
\end{align}
where $\bar{\pmb v}_k^H \triangleq \pmb v_k^H\pmb\Lambda_k^{1/2}$ is the equivalent channel between the BS and the $k$-th user.
A detailed block diagram of the proposed JSDD scheme is depicted in Fig. \ref{systemModel}. \footnote{
We note that the proposed JSDD scheme can be extended to the cases where the BS are equipped with an uniform planar array. More specifically, as shown in \cite{covMatForm}, for a rectangular antenna array with $N$ rows and $M$ columns, the channel covariance matrix $\pmb{R}_k$ can be written as $\pmb{R}_k = \pmb{R}_{k,H}\otimes\pmb{R}_{k,V}$, where $\pmb{R}_{k,H}\in\mathbb{C}^{M\times M}$ and $\pmb{R}_{k,V}\in\mathbb{C}^{N\times N}$ are the horizontal and vertical
channel covariance matrices, respectively. Note that as $M$ ($N$) becomes large, the eigenvectors of $\pmb{R}_{k,H}$ ($\pmb{R}_{k,V}$) can be approximated by selecting the columns from a DFT-matrix according to Proposition 1. As a results, the beamforming design in the case of UPA follows the same as that will be discussed in this paper.
}

Now each user has an equivalent channel vector $\bar{\pmb v}_k$ with  the covariance matrix $\bar{\pmb R}_{kk} = \mathbb{E}\left\{\bar{\pmb v}_k\bar{\pmb v}_k^H\right\}=\pmb\Lambda_k$.
The matrix $\pmb M_k$ becomes a precoding matrix after the OSTBC code matrix of user $k$.
 Assume that the estimated equivalent channel is $\hat{\bar{\pmb v}}_k$, and the true channel and estimated channel have the relation
\begin{align}
\bar{\pmb v}_k = \xi_k\hat{\bar{\pmb v}}_k+\sqrt{1-\xi_k^2}\pmb\tau_k,\label{channelEstModel}
\end{align}
where $\bar{\pmb v}_k$ and $\hat{\bar{\pmb v}}_k$ are jointly complex Gaussian distributed with correlation coefficient $\xi_k$, and $\pmb\tau_k$ is the i.i.d. complex Gaussian estimation error vector with zero mean and variance matrix $\pmb\Lambda_k$.
Obviously, in \eqref{channelEstModel}, if $\xi_k=1$, then we have $\bar{\pmb v}_k = \hat{\bar{\pmb v}}_k$, which means that the BS has the perfect knowledge of the $k$-th user's CSI. However, in this paper, we only focus on the cases where $\xi_k<1$ for $k=1,2,\cdots,K$, i.e., the CSITs are imperfect.

Denote the total transmit power of the BS as $P_T$. In general, $P_T$ is limited by the maximal power budget of the BS, denoted by $P_{T,\max}$, which can be written as
\begin{align}
P_T &= \sum_{k=1}^K P_k = \sum_{k=1}^K \frac{1}{T}\mathbb{E}\left\{\left\|\pmb{W}_k\pmb{Z}_k\right\|_F^2\right\} \nonumber\\
&=
\sum_{k=1}^K\frac{L_k}{T}\left\|\pmb{W}_k\right\|_F^2
=\sum_{k=1}^K\frac{L_k}{T}\left\|\pmb{M}_k\right\|_F^2\leq P_{T,\max}.\label{PowerExpression}
\end{align}

\subsection{Pairwise error probability analysis}
In this subsection we analyze the PEP of each user under the channel model and the proposed JSDD scheme. Since for OSTBC the maximal likelihood (ML) detection is equivalent to symbol-by-symbol detection, we assume each user uses ML detection.
It is also reasonable to assume that the BS has partial information about the equivalent downlink channel, i.e., $\hat{\bar{\pmb v}}_k$, while the $k$-th user has the perfect information about the equivalent downlink channel, i.e., $\bar{\pmb v}_k$.

For user $k$, let  $\bar{\pmb Z}_k\triangleq\pmb M_k\pmb Z_k$. With the ML decoder, the decoding criteria is
\begin{align}
\hat{\bar{\pmb Z}}_k = \mathrm{arg} \min_{\bar{\pmb Z}_k} \left\|\pmb y_k-\bar{\pmb v}_k^H\bar{\pmb Z}_k\right\|_F,
\end{align}
where $\hat{\bar{\pmb Z}}_k$ is the codeword decision. Computing the codeword error probability directly is difficult, if not impossible. Similar to \cite{STBCAndBeamforming}, we consider PEP conditioned on the estimated channel $\hat {\bar{\pmb v}}_k^H$, which is given by \eqref{intExpression},
\begin{figure*}[t]
\begin{align}
&\mathcal{P}\left(\bar{\pmb Z}_k\rightarrow \hat{\bar{\pmb Z}}_k|\hat{\bar{\pmb v}}_k^H\right)=\int{\mathcal{P}\left(\bar{\pmb Z}_k\rightarrow\hat{\bar{\pmb Z}}_k|\bar{\pmb v}_k^H,\hat{\bar{\pmb v}}_k^H\right)p(\bar{\pmb v}_k^H|\hat{\bar{\pmb v}}_k^H)d\bar{\pmb v}_k^H} \nonumber\\
=&\int{\mathcal{Q}\left(\sqrt{\frac{\left\|\bar{\pmb v}_k^H\left(\bar{\pmb Z}_k-\hat{\bar{\pmb Z}}_k\right)\right\|_F^2}{2\sigma^2}}\right)p(\bar{\pmb v}_k^H|\hat{\bar{\pmb v}}_k^H)d\bar{\pmb v}_k^H}
\leq \int{\frac{1}{2}\exp\left(-\frac{\left\|\bar{\pmb v}_k^H\left(\bar{\pmb Z}_k-\hat{\bar{\pmb Z}}_k\right)\right\|^2}{4\sigma^2}\right)p(\bar{\pmb v}_k^H|\hat{\bar{\pmb v}}_k^H)d\bar{\pmb v}_k^H}.\label{intExpression}
\end{align}
\rule{\textwidth}{.2mm}
\vspace{-7mm}
\end{figure*}
where $\mathcal{Q}(x)$ is the Gaussian tail function. In \eqref{intExpression}, the first equality follows the total probability law and the last inequality is the Chernoff bound. According to \eqref{channelEstModel}, for $\xi_k < 1$, we have
\begin{align}
p(\bar{\pmb v}_k^H|\hat{\bar{\pmb v}}_k^H) = \frac{\exp{\left(- \frac{1}{(1-\xi_k^2)}\left(\bar{\pmb v}_k^H - \xi_k\hat{\bar{\pmb v}}_k^H\right)\pmb \Lambda_k^{-1}\left(\bar{\pmb v}_k-\xi_k\hat{\bar{\pmb v}}_k\right)\right)}}{\pi^{r_k}\det\left((1-\xi_k^2)\pmb\Lambda_k\right)},\nonumber
\end{align}
and substitute $p(\bar{\pmb v}_k^H|\hat{\bar{\pmb v}}_k^H)$ into \eqref{intExpression}, the exponent part of the integrand can be written as \eqref{numerator}
\begin{figure*}[t]
\begin{align}
&-\frac{||\bar{\pmb v}_k^H\left(\bar{\pmb Z}_k-\hat{\bar{\pmb Z}}_k\right)||_F^2}{4\sigma^2}-\left(\bar{\pmb v}_k^H - \xi_k\hat{\bar{\pmb v}}_k^H\right)\pmb{A}_k^{-1}\left(\bar{\pmb v}_k-\xi_k\hat{\bar{\pmb v}}_k\right)- \left(\bar{\pmb v}_k -  \xi_k\pmb B_k^{-1}\pmb{A}_k^{-1}\hat{\bar{\pmb v}}_k \right)^H\pmb B_k\left(\bar{\pmb v}_k -  \xi_k\pmb B_k^{-1}\pmb{A}_k^{-1}\hat{\bar{\pmb v}}_k \right)\nonumber\\
=&\ \pmb\mu_k^H\left(\pmb B_k^{-1}- \pmb{A}_k\right)\pmb\mu_k - \left(\bar{\pmb v}_k-\pmb B_k^{-1}\pmb\mu_k\right)^H\pmb B_k\left(\bar{\pmb v}_k-\pmb B_k^{-1}\pmb\mu_k\right),\label{numerator}
\end{align}
\rule{\textwidth}{.2mm}
\vspace{-7mm}
\end{figure*}
where
$\pmb{A}_k \triangleq \left(1 - \xi_k^2\right)\pmb{\Lambda}_k$,
$\pmb \mu_k\triangleq \xi_k\pmb{A}_k^{-1}\hat{\bar{\pmb v}}_k$ and
$\pmb{B}_k \triangleq \frac{1}{4\sigma^2}\left(\bar{\pmb Z}_k-\hat{\bar{\pmb Z}}_k\right)\left(\bar{\pmb Z}_k-\hat{\bar{\pmb Z}}_k\right)^H+\pmb{A}_k^{-1}$.
Hence, combing $p(\bar{\pmb v}_k^H|\hat{\bar{\pmb v}}_k^H)$ and \eqref{numerator}, for $\xi_k<1$, the conditioned PEP \eqref{intExpression} is upper bounded by \eqref{intResult}.
\begin{figure*}[t]
\begin{align}
\mathcal{P}(\bar{\pmb Z}_k\rightarrow\hat{\bar{\pmb Z}}_k|\hat{\bar{\pmb v}}_k^H) & \leq\frac{\det\left(\pmb B_k^{-1}\right)\exp\left(\pmb\mu_k^H\left(\pmb B_k^{-1}-\pmb{A}_k\right)\pmb\mu_k\right)}{2\det\left(\pmb{A}_k\right)}\int{\frac{\exp\left(-\left(\bar{\pmb v}_k^H-\pmb\mu_k^H\pmb B_k^{-1}\right)\pmb B_k\left(\bar{\pmb v}_k-\pmb B_k^{-1}\pmb\mu_k\right)\right)}{\pi^{r_k}\det\left(\pmb B_k^{-1}\right)}}d\bar{\pmb v}_k^H\nonumber\\
&=g_k(\pmb M_k)\triangleq\frac{\det\left(\pmb B_k^{-1}\right)\exp\left(\pmb\mu_k^H\left(\pmb B_k^{-1}-\pmb{A}_k\right)\pmb\mu_k\right)}{2\det\left(\pmb{A}_k\right)}.\label{intResult}
\end{align}
\rule{\textwidth}{.2mm}
\vspace{-7mm}
\end{figure*}
The last equality in \eqref{intResult} holds due to the integral term equals one since it is a integral of a complex Gaussian pdf.
Since $\bar{\pmb Z}_k=\pmb M_k\pmb Z_k$  and $\pmb Z_k$ is an OSTBC block, based on  \eqref{OSTBCForm}, $\pmb B_k$ could be rewritten as
\begin{align}
\pmb B_k = \rho_{\pmb s_k,\hat{\pmb s}_k}\pmb M_k\pmb M_k^H+\pmb{A}_k^{-1}, \label{tempVar}
\end{align}
where $
\rho_{\pmb s_k,\hat{\pmb s}_k}\triangleq \frac{||\pmb s_k-\hat{\pmb s}_k||_2^2}{4\sigma^2}$,
and $\pmb s_k$ and $\hat{\pmb s}_k$ are the symbol vectors corresponding to the codeword matrices $\bar{\pmb Z}_k$ and $\hat{\bar{\pmb Z}}_k$, respectively.
We note that $g_k(\pmb M_k)$ is related to $\rho_{\pmb s_k,\hat{\pmb s}_k}$, which is determined by the symbol vector pair $\left(\pmb s_k,\hat{\pmb s}_k\right)$. And it is not difficult to verify that $g_k(\pmb M_k)$ is a decreasing function of $\rho_{\pmb s_k,\hat{\pmb s}_k}$, which means that the error probability is dominated by the codeword pairs with minimal value of $\rho_{\pmb s_k,\hat{\pmb s}_k}$. Hence, we should design $\pmb M_k$ to minimize
\begin{align}
\tilde{g}_k(\pmb M_k)\triangleq
g_k(\pmb M_k) |_{\rho_{\pmb s_k,\hat{\pmb s}_k}=\rho_{k,\min}}, \label{OF}
\end{align}
where $\rho_{k,\min} = \min_{\pmb s_k \neq \hat{\pmb s}_k } \rho_{\pmb s_k,\hat{\pmb s}_k}$.
For example, recall that we have assumed that $\mathbb{E}|s_{k,i}|^2=1$, and therefore,  for QPSK modulation, $\rho_{k,\min}=\frac{2}{4\sigma^2} =\frac{1}{2\sigma^2}$, and for BPSK modulation, $\rho_{k,\min} =\frac{4}{4\sigma^2} =\frac{1}{\sigma^2}$.

So far, we have shown that for each user, the maximal PEP of the ML decoding is upper bounded by \eqref{OF}, and
the problem now becomes how to  design $\pmb M_k, k=1,2, \cdots, K$, to minimize \eqref{OF} under a pre-given transmit power constraint.
{ Basically, the sum power constraint of all $K$ users should be considered since they are all transmitted from a same BS.
However, from (\ref{partialCSIChannel}) we know that each user could be handled independently if the constraint $P_T\leq P_{T,\max}$ is decomposed into $K$ independent constraints on $P_k$ for $1\leq k \leq K$, i.e., $P_k\leq P_{k,\max}$ with $P_{k,\max}$ being the maximum power of the $k\text{-th}$ user. We note that it is useful to consider the individual power constraint for the following reason:
\begin{enumerate}
\item Under the individual user power constraint, the PEP minimization problem for multiple users can be divided into several parallel sub-problems. Solving the sub-problems separately can be computationally efficient, and for some special cases, the optimal closed-from solutions can be obtained as shown in Section IV.
\item Simulation results show that uniformly allocating the power for the users does not cause too much performance loss, especially when the power budget of the BS is low as shown in Section VI.
    Therefore, the individual user power constraint can be viewed as a trade-off between the computational complexity and the achievable system performance.
\item Solving the individual user power constrained PEP minimization problem provides a guidance for solving the PEP minimization problem under the sum power constraint. In fact, the proposed numerical algorithm for solving the latter problem is partially based on that for the former one.
\end{enumerate}
Based on the above observations, in the following, the individual user power constraint is discussed first, and after that we consider the sum power constraint.}

\section{Individual User Power Constraint\label{indSection}}
In this section, we consider the individual user power constraint, namely,
$P_k\leq P_{k,\max}$ for $k=1,2,\cdots,K$ with $P_{k,\max}$ being a constant, and $\sum_{k=1}^KP_{k,\max} = P_{T,\max}$.
These power constraints are equivalent to $\left\|\pmb M_k\right\|_F^2\leq\alpha_k, k=1,2,\cdots,K$ with $\alpha_k\triangleq \frac{TP_{k,\max}}{L_k}$.
According to \eqref{partialCSIChannel}, we can design $\pmb M_k$ for each user independently.

According to the PEP criteria, our goal is to design $\pmb M_k$ to minimize the objective function in \eqref{OF}, and therefore, the optimization problem can be written as
\begin{align}
\label{optProblem}
\min_{\pmb M_k}&\quad \tilde f_k(\pmb M_k),\quad \mathrm{s.t.}~\left\|\pmb M_k\right\|_F^2\leq\alpha_k,
\end{align}
where $\tilde{f}_k\left(\pmb M_k\right) \triangleq \ln \left(\tilde{g}_k\left(\pmb M_k\right)\right) =
\pmb\mu_k^H (\rho_{k,\min}\pmb M_k\pmb M_k^H+\pmb{A}_k^{-1} )^{-1}\pmb\mu_k - \ln\det (\rho_{k,\min}\pmb M_k\pmb M_k^H+\pmb{A}_k^{-1} )-b_k$ with $b_k$ being a constant given by $b_k\triangleq \pmb{\mu}_k^H\pmb{A}_k\pmb{\mu}_k+\ln2+\ln\det\left(\pmb{A}_k\right)$.

Optimizing problem \eqref{optProblem} directly is difficult since the object function $\tilde f_k(\pmb M_k)$ is non-convex. In the following we focus on solving this problem. We first provide an SDR-based method and a sufficient condition to achieve the global optimum.
We then propose a low-complexity SCA-based iterative optimization method to obtain a suboptimal solution.
Furthermore, some special cases with closed-form solutions are also discussed, including the hign SNR, low SNR and no effective CSIT cases.

\subsection{SDR solution}
\label{SDRIndividualPower}
In order to handle the problem, we first let $\pmb\Omega_k=\pmb M_k\pmb M_k^H$. Then, the optimization problem becomes
\begin{align}
\label{finalOptProblem}
\min_{\pmb\Omega_k,\mathrm{rank}(\pmb\Omega_k)\leq N_k}&\quad \breve{f}_k(\pmb\Omega_k),\\
\mathrm{s.t.}\quad\quad\  & \quad \mathrm{tr}(\pmb\Omega_k)\leq\alpha_k,\ \pmb\Omega_k = \pmb\Omega_k^H\succeq0,
\end{align}
with
$
\breve{f}_k(\pmb\Omega_k)\triangleq\pmb\mu_k^H(\rho_{k,\min}\pmb\Omega_k+\pmb{A}_k^{-1})^{-1}\pmb\mu_k
-\ln\det(\rho_{k,\min}\pmb\Omega_k+\pmb{A}_k^{-1}). \nonumber
$
The rank constraint is due to $\pmb M_k\in\mathbb{C}^{r_k\times N_k}$ and $N_k\leq r_k$.
We know that the rank constraint is nonconvex. To deal with it, SDR method is used, i.e., we remove the rank constraint.

\subsubsection{SDR method}
Without the rank constraint, the problem (\ref{finalOptProblem}) is convex. To see it clearly, by introducing a slack variable $\eta$, \eqref{finalOptProblem} is equivalent to
\begin{subequations}
\label{equiProblem}
\begin{align}
\min_{\pmb\Omega_k,\eta}&\quad \eta-\ln\det\left(\rho_{k,\min}\pmb\Omega_k+
\pmb{A}_k^{-1}\right)\\
\mathrm{s.t.}&\quad\pmb\mu_k^H\left(\rho_{k,\min}\pmb\Omega_k+
\pmb{A}_k^{-1}\right)^{-1}\pmb\mu_k\leq \eta,\label{equiProblemFirst}\\
&\quad \mathrm{tr}(\pmb\Omega_k)\leq\alpha_k,\ \pmb\Omega_k = \pmb\Omega_k^H\succeq0,
\end{align}
\end{subequations}
due to the fact that the inequality constraint in \eqref{equiProblemFirst} is active at the optimum. In \eqref{equiProblem}, it is obvious that the objective function is convex, and by using Schur complement, \eqref{equiProblemFirst} is equivalent to
\begin{align}
  \begin{bmatrix}
    \rho_{k,\min}\pmb\Omega_k+\pmb{A}_k^{-1}& \pmb\mu_k \\
    \pmb\mu_k^H & \eta \\
  \end{bmatrix}
\succeq 0, \label{equiConstraint}
\end{align}
which is also convex. Therefore, we conclude that (\ref{equiProblem})  is a convex optimization problem \cite{convexOpt}.

Now the problem \eqref{equiProblem} is an SDP problem, which can be solved efficiently \cite{cvxToolbox}. If the optimal solution $\pmb\Omega_k^o$ of \eqref{equiProblem} satisfies the rank constraint of (\ref{finalOptProblem}), then eigenvalue decomposition is used to obtain the optimal $\pmb M_k^o$. Otherwise,  Gaussian randomization  method could be used to obtain a suboptimal solution.

Since whether the optimal $\pmb M_k^o$ could be obtained depends on the rank of $\pmb\Omega_k^o$, we next give a sufficient condition to guarantee that the rank constraint is met, i.e., the global optimum of (\ref{optProblem}) is obtained.

\subsubsection{A sufficient condition achieving global optimum}
The Lagrange function of (\ref{equiProblem}) is
\begin{align}
&L(\pmb\Omega_k,\eta,\kappa_1,\kappa_2,\pmb Q) \nonumber \\
=& \eta-\ln\det\left(\rho_{k,\min}\pmb\Omega_k+\pmb{A}_k^{-1}\right)+
\kappa_2\left(\mathrm{tr}(\pmb\Omega_k)-\alpha_k\right) \nonumber\\
& -\mathrm{tr}(\pmb Q\pmb\Omega_k) + \kappa_1\left(\pmb\mu_k^H\left(\rho_{k,\min}\pmb\Omega_k+
\pmb{A}_k^{-1}\right)^{-1}\pmb\mu_k-\eta\right),
\end{align}
where $\kappa_1, \kappa_2 \geq0$ and $\pmb Q\succeq0$ are the lagrange multipliers. Define $\tilde{\pmb B}_k(\pmb\Omega_k)\triangleq\left(\rho_{k,\min}\pmb\Omega_k+\pmb{A}_k^{-1}\right)^{-1}$, and we have
the following proposition.
\begin{prop}
If $
\mathrm{rank}\left(\kappa_2\pmb I-\rho_{k,\min}\tilde{\pmb B}_k^{T}(\pmb\Omega_k)\right)\geq r_k+1- N_k$,
we have $\mathrm{rank}(\pmb\Omega_k)\leq N_k$, i.e., the rank constraint is satisfied.
\end{prop}
\begin{IEEEproof}
Please refer to Appendix B.
\end{IEEEproof}
In fact, the SDR method provides a lower bound on the object function of problem \eqref{optProblem}, and establishes a performance benchmark. In the following, we propose an SCA-based iterative method to solve the original problem \eqref{optProblem}.

\subsection{SCA Method}
\label{SCAIndividualPower}
The basic idea of SCA method is to approximate the original non-convex constraint or objective function with a convex function that satisfies some certain properties, and solve the so-obtained approximate convex problem iteratively \cite{SCA1,SCA2}. The iteration can be guaranteed to converge to a stationary solution of the original non-convex  problem.

To use the SCA method to solve the problem (\ref{optProblem}), we first rewrite the objective function as
\begin{align}
\tilde{f}_k\left(\pmb M_k\right)=&\ \pmb\omega_k^H\left(\pmb C_k \pmb {M}_k\pmb M_k^H\pmb C_k^H+\pmb I\right)^{-1}\pmb\omega_k  \nonumber \\ & -\ln\det\left(\pmb C_k\pmb{M}_k\pmb M_k^H\pmb C_k^H+\pmb I\right)-\tilde{b}_k,\nonumber
\end{align}
where $\tilde{b}_k \triangleq b_k + \ln\det\left(\pmb A_k^{-1}\right)$, $\pmb\omega_k\triangleq\pmb A_k^{1/2}\pmb\mu_k$, and $\pmb C_k\triangleq\sqrt{\rho_{k,\mathrm{min}}}\pmb A_k^{1/2}$. Using the matrix inversion lemma, we have
\begin{align}
& \pmb\omega_k^H\left(\pmb C_k\pmb{M}_k\pmb M_k^H\pmb C_k^H+\pmb I\right)^{-1}\pmb\omega_k \nonumber \\
=&\pmb\omega_k^H\left(\pmb I-\pmb C_k\pmb{M}_k\left(\pmb M_k^H\pmb C_k^H\pmb C_k\pmb{M}_k+\pmb I\right)^{-1}\pmb M_k^H\pmb C_k^H\right)\pmb\omega_k\nonumber\\
=&\pmb\omega_k^H\pmb\omega_k-\pmb\omega_k^H\pmb C_k\pmb{M}_k\pmb G_k^{-1}\pmb M_k^H\pmb C_k^H\pmb\omega_k, \nonumber
\end{align}
where $\pmb G_k\triangleq\pmb M_k^H\pmb C_k^H\pmb C_k\pmb{M}_k+\pmb I$. Similarly, the second term of $f_k(\pmb M_k)$ can be rewritten as
\begin{align}
\begin{aligned}
&-\ln\det\left(\pmb C_k\pmb{M}_k\pmb M_k^H\pmb C_k^H+\pmb I\right) \nonumber \\
=&\ln\det\left(\pmb I-\pmb C_k\pmb{M}_k\left(\pmb M_k^H\pmb C_k^H\pmb C_k\pmb{M}_k+\pmb I\right)^{-1}\pmb M_k^H\pmb C_k^H\right)\nonumber \\
=&\ln\det(\pmb P_k), \nonumber
\end{aligned}
\end{align}
where $\pmb P_k\triangleq\pmb I-\pmb C_k\pmb{M}_k\left(\pmb M_k^H\pmb C_k^H\pmb C_k\pmb{M}_k+\pmb I\right)^{-1}\pmb M_k^H\pmb C_k^H$. Now, the optimization problem (\ref{optProblem}) is equivalent to
\begin{align}
\min_{\pmb M_k}\  \tilde{f}_{k,1}\left(\pmb M_k\right)+\tilde{f}_{k,2}\left(\pmb M_k\right) + c_k,~\mathrm{s.t.}\ \left\|\pmb M_k\right\|_F^2\leq\alpha_k, \label{SCAIgnoreConstant}
\end{align}
where $\tilde{f}_{k,1}\left(\pmb M_k\right)\triangleq-\pmb\omega_k^H\pmb C_k\pmb{M}_k\pmb G_k^{-1}\pmb M_k^H\pmb C_k^H\pmb\omega_k$, $\tilde{f}_{k,2}\left(\pmb M_k\right)\triangleq\ln\det(\pmb P_k)$, and $c_k = \pmb\omega_k^H\pmb\omega_k - \tilde{b}_k$ is a constant.

Both $\tilde{f}_{k,1}\left(\pmb M_k\right)$ and $\tilde{f}_{k,2}\left(\pmb M_k\right)$ are non-convex, thus we now reform problem (\ref{SCAIgnoreConstant}) using SCA principle. The basic idea of the SCA method to solve \eqref{SCAIgnoreConstant} iteratively is to find the upper convex approximate function of the objective function at the $i$-th iteration, and then minimize the so-obtained convex objective function. The acquired optimal solution is then used to construct a new convex approximate objective function which will be minimized in the $(i+1)$-th iteration.

To solve \eqref{SCAIgnoreConstant}, upper convex approximations for both $\tilde{f}_{k,1}\left(\pmb M_k\right)$ and $\tilde{f}_{k,1}\left(\pmb M_k\right)$ are provided in the following proposition to enable the SCA iteration.

\begin{prop}
For arbitrary constant matrix $\hat{\pmb{M}}_{k,i}$, upper convex approximations of $\tilde{f}_{k,j}\left(\pmb M_k\right)$ for $j \in \{1,2\}$ are given by
\begin{align}
\tilde{f}_{k,j}\left(\pmb M_k\right)
&\leq\hat{f}_{k,j}\left(\pmb M_k,\hat{\pmb M}_{k,i}\right),\quad j \in \{1,2\}, \label{firstUpperBound}
\end{align}
where
$
\hat{f}_{k,1}(\pmb M_k,\hat{\pmb M}_{k,i})
\triangleq
\mathrm{tr}(\pmb M_{k}^H\pmb C_k^H\pmb C_k\pmb{M}_{k}\pmb{\gamma}_{k,i}\pmb{\gamma}_{k,i}^H ) -2\mathrm{Re}(\mathrm{tr}(\pmb{M}_k^H\pmb C_k^H\pmb\omega_k\pmb\omega_k^H\pmb C_k
\hat{\pmb M}_{k,i}\hat{\pmb G}_{k,i}^{-1}))
+a_{k,i}^{(1)}$,
$\hat{f}_{k,2}(\pmb M_k,\hat{\pmb M}_{k,i})
\triangleq
\mathrm{tr}(\pmb{M}_{k}^H\pmb C_k^H\pmb C_k\pmb{M}_{k}\pmb{\Xi}_{k,i}) -
2\mathrm{Re}(\mathrm{tr}(\pmb{M}_k^H\pmb C_k^H\hat{\pmb P}_{k,i}^{-1}\pmb C_k\hat{\pmb M}_{k,i}\hat{\pmb G}_{k,i}^{-1})) + a_{k,i}^{(2)}$,
$\pmb{\gamma}_{k,i} = \hat{\pmb G}_{k,i}^{-1}\hat{\pmb M}_{k,i}^H\pmb C_k^H\pmb{\omega}_k$,
$\hat{\pmb G}_{k,i}\triangleq \hat{\pmb M}_{k,i}^H\pmb C_k^H\pmb C_k\hat{\pmb{M}}_{k,i}+\pmb I$,
$\pmb{\Xi}_{k,i} = \hat{\pmb G}_{k,i}^{-1}\hat{\pmb M}_{k,i}^H\pmb C_k^H\hat{\pmb P}_{k,i}^{-1}\pmb C_k\hat{\pmb{M}}_{k,i}\hat{\pmb G}_{k,i}^{-1}$,
$\hat{\pmb P}_{k,i}\triangleq (\pmb I + \pmb C_k\hat{\pmb{M}}_{k,i}\hat{\pmb M}_{k,i}^H\pmb C_k^H)^{-1}$,
$a_{k,i}^{(1)} = \pmb{\gamma}_{k,i}^H\pmb{\gamma}_{k,i}$, and
$a_{k,i}^{(2)} = \mathrm{tr}(\pmb{\Xi}_{k,i}) +\ln\det(\hat{\pmb P}_{k,i})+\mathrm{tr}(\hat{\pmb P}_{k,i}^{-1}-I)$.

\end{prop}
\begin{IEEEproof}
The detailed derivations of \eqref{firstUpperBound} are provided in the Appendix C.
\end{IEEEproof}

It is not hard to see that both $\hat{f}_{k,1}(\pmb M_k,\hat{\pmb M}_{k,i})$ and $\hat{f}_{k,2}(\pmb M_k,\hat{\pmb M}_{k,i})$ are  convex functions of $\pmb M_k$. Then, by ignoring the constant terms, in the $i$-th iteration, the SCA optimization problem is
\begin{align}
&\min_{\left\|\pmb M_k\right\|_F^2\leq\alpha_k}\ \hat{f}_{k}^{(i)}\left(\pmb M_k\right)\nonumber \\
\Rightarrow&
\min_{\left\|\pmb M_k\right\|_F^2\leq\alpha_k}\
\left\{\begin{aligned}
\mathrm{tr}
&\left(\pmb M_{k}^H\pmb C_k^H\pmb C_k\pmb{M}_{k}\left(\pmb{\gamma}_{k,i}\pmb{\gamma}_{k,i}^H + \pmb{\Xi}_{k,i}\right)\right)\\
&-2\mathrm{Re}\left(\mathrm{tr}\left(\pmb{M}_k^H\pmb{\Gamma}_{k,i}\right)\right)
\end{aligned}\right\}
,\label{SCAConvexProblem}
\end{align}
where $\hat{f}_{k}^{(i)}\left(\pmb M_k\right) \triangleq \hat{f}_{k,1}(\pmb M_k,\hat{\pmb M}_{k,i})+\hat{f}_{k,2}(\pmb M_k,\hat{\pmb M}_{k,i})$ and $\pmb{\Gamma}_{k,i}\triangleq\pmb C_k^H\left(\pmb\omega_k\pmb\omega_k^H+\hat{\pmb P}_{k,i}^{-1}\right)\pmb C_k
\hat{\pmb M}_{k,i}\hat{\pmb G}_{k,i}^{-1}$.
Let the solution of (\ref{SCAConvexProblem}) be $\pmb M_{k,i}^o$. In the $(i+1)$-th iteration, we have $\hat{\pmb{M}}_{k,i+1}=\pmb M_{k,i}^o$, where $\hat{\pmb{M}}_{k,i+1}$ is the constant matrix in the $(i+1)$-th iteration. In such a way, we solve the problem (\ref{SCAConvexProblem}) iteratively until convergence or the maximum iteration number is reached.
We have to point out that \eqref{SCAConvexProblem} is in fact a QCQP-1 problem, which can be simply solve by bisection method as shown in \cite{K.Huang2016}.
For the initialization step of the SCA method, we can randomly select a matrix $\pmb M_{k,1}$ satisfying the constraint $\left\|\pmb M_{k,1}\right\|_F^2\leq\alpha_k$.

Both the SDR and SCA method could not guarantee the optimal solution. In the following, we will consider some special cases where the optimal solutions are obtained.

\subsection{Special cases}
\subsubsection{High SNR}
In the high SNR regime, we have $\sigma^2 \rightarrow 0$ and $\rho_{k,\min}\rightarrow+\infty$, and thus, we can get
\begin{align}
\tilde{f}_k\left(\pmb M_k\right)
&= \pmb\omega_k^H\left(\rho_{k,\min}\pmb{A}_k^{\frac{1}{2}} \pmb {M}_k\pmb M_k^H\pmb{A}_k^{\frac{1}{2}}+\pmb{I}_{r_k}\right)^{-1}\pmb\omega_k \nonumber \\
&\quad -\ln\det\left(\rho_{k,\min}\pmb{A}_k^{\frac{1}{2}}\pmb{M}_k\pmb M_k^H\pmb{A}_k^{\frac{1}{2}}+\pmb{I}_{r_k}\right)-\tilde{b}_k,\nonumber\\
&\approx -\ln\det\left(\pmb{I}_{N_k} + \rho_{k,\min}\pmb M_k^H\pmb{A}_k\pmb{M}_k\right),\label{HSNR}
\end{align}
where \eqref{HSNR} is because $\det(\pmb{I} + \pmb{AB})=\det(\pmb{I} + \pmb{BA})$, and as $\rho_{k,\min}\rightarrow+\infty$, we have
$\pmb\omega_k^H(\rho_{k,\min}\pmb{A}_k^{\frac{1}{2}} \pmb {M}_k\pmb M_k^H\pmb{A}_k^{\frac{1}{2}}+\pmb{I}_{r_k})^{-1}\pmb\omega_k = 0$ if $\mathrm{rank}(\pmb {M}_k\pmb M_k^H)=r_k$ and $\pmb\omega_k^H(\rho_{k,\min}\pmb{A}_k^{\frac{1}{2}} \pmb {M}_k\pmb M_k^H\pmb{A}_k^{\frac{1}{2}}+\pmb{I}_{r_k})^{-1}\pmb\omega_k = O(1)$ if $\mathrm{rank}(\pmb {M}_k\pmb M_k^H)<r_k$,
where $O(1)$ denotes a constant that does not depend on $\rho_{k,\min}$. By ignoring the constant terms, the optimization problem \eqref{optProblem} becomes
\begin{align}\label{infSNR}
\max_{\left\|\pmb M_k\right\|_F^2\leq\alpha_k} \tilde{f}_k^{(\mathrm{H})}\left(\pmb M_k\right),
\end{align}
where $\tilde{f}_k^{(\mathrm{H})}(\pmb M_k)\triangleq\ln\det(\pmb I + \rho_{k,\min}\pmb M_k^H\pmb{A}_k\pmb{M}_k)$.
Based on \eqref{infSNR}, the following proposition characterizes the optimal beamforming matrix, denoted by $\pmb M_{k,\mathrm{opt}}$, at high SNR region.
\begin{prop}
\label{P4}
At high SNR region, $\pmb M_{k,\mathrm{opt}}$ is given by
\begin{align}
\pmb M_k^o=\left[\sqrt{\frac{\alpha_k}{N_k}}\pmb{I}_{N_k\times N_k},\pmb{0}_{N_k\times (r_k-N_k)}\right]^T.
\end{align}
\end{prop}
\begin{IEEEproof}
Denote the singular value decomposition of $\pmb{M}_{k}$ as $\pmb{M}_{k} = \pmb{U}_{\pmb{M}_{k}}\pmb{D}_{\pmb{M}_{k}}\pmb{V}_{\pmb{M}_{k}}^H$, where $\pmb{U}_{\pmb{M}_{k}}\in\mathbb{C}^{r_k\times N_k}$ is a sub-unitary matrix, $\pmb{D}_{\pmb{M}_{k}} = \mathrm{diag}\left(\sqrt{q_{k,1}},\sqrt{q_{k,2}},\cdots,\sqrt{q_{k,N_k}}\right)$, and $\pmb{V}_{\pmb{M}_{k}}\in\mathbb{C}^{N_k\times N_k}$ is an unitary matrix. Then, $\tilde{f}_k^{(\mathrm{H})}\left(\pmb M_k\right)$ in \eqref{infSNR} can be written as
\begin{align}
\tilde{f}_k^{(\mathrm{H})}\left(\pmb M_k\right) = \ln\det \left(\pmb I + \rho_{k,\min} \pmb{D}_{\pmb{M}_{k}}\pmb{U}_{\pmb{M}_{k}}^H\pmb{A}_k\pmb{U}_{\pmb{M}_{k}}\pmb{D}_{\pmb{M}_{k}}\right). \label{HSOF}
\end{align}
Note that \eqref{HSOF} does not depend on $\pmb{V}_{\pmb{M}_k}$, and therefore, we can simply set $\pmb{V}_{\pmb{M}_k}=\pmb{I}_{N_k}$. Due to the fact that $\pmb{D}_{\pmb{M}_{k}}$ is a diagonal matrix, \eqref{HSOF} is maximized when $\pmb{U}_{\pmb{M}_{k}}^H\pmb{A}_k\pmb{U}_{\pmb{M}_{k}}$ is diagonal. Recall that $\pmb{A}_k = \left(1 - \xi_k^2\right)\pmb{\Lambda}_k$ is already a diagonal matrix with its diagonal elemets being in decreasing order, so we have $\pmb{U}_{\pmb{M}_{k}} = \left[\pmb{I}_{N_k},\pmb{0}_{(r_k - N_k)\times N_k}\right]^T$. As a result, problem \eqref{infSNR} becomes,
\begin{align}\label{infSNRS2}
\begin{aligned}
&\max\ \quad  \sum_{i=1}^{N_k} \ln(1 + \left(1 - \xi_k^2\right)\rho_{k,\min}\lambda_{k,i}q_{k,i} ),\\
&\ \ \mathrm{s.t.}\ \quad \sum_{i=1}^{N_k} q_{k,i} \leq \alpha_k.
\end{aligned}
\end{align}
The Lagrange function of \eqref{infSNRS2} is
\begin{align}
L(q_{k,i},\kappa) &= \sum_{i=1}^{ N_k}\ln(1 + \left(1 - \xi_k^2\right)\rho_{k,\min}\lambda_{k,i}q_{k,i} ) \nonumber \\
&\quad  - \kappa\left(\sum_{i=1}^{ N_k}q_{k,i}-\alpha_k\right),\nonumber
\end{align}
where $\kappa> 0$ is the Lagrange multiplier. The corresponding KKT conditions are
\begin{align}
& \frac{\partial L(q_{k,i},\kappa)}{\partial q_{k,i}} = \frac{\left(1 - \xi_k^2\right)\rho_{k,\min}\lambda_{k,i}}{1+\left(1 - \xi_k^2\right)\rho_{k,\min}\lambda_{k,i}q_{k,i}}-\kappa = 0,\\
& \kappa\left(\sum_{i=1}^{ N_k}q_{k,i}-\alpha_k\right) = 0.
\end{align}
Then the solution to \eqref{infSNR} has the form of waterfilling as
\begin{align}
q_{k,i} = \left[\frac{1}{\kappa}-\frac{1}{\left(1 - \xi_k^2\right)\rho_{k,\min}\lambda_{k,i}}\right]^+,\label{waterfillingsolution}
\end{align}
where $[x]^+=\max(x,0)$ and $\kappa$ is selected to satisfy $\sum_{i=1}^{ N_k}q_{k,i}=\alpha_k$. Note that as $\rho_{k,\min}\rightarrow+\infty$, we obtain $q_{k,1}=q_{k,2}=\cdots=q_{k,N_k}=\frac{1}{\kappa}=\frac{\alpha_k}{N_k}$.
\end{IEEEproof}
According to Proposition 4, at the high SNR region, the transmit power is uniformly allocated to the $N_k$ OSTBC streams.
Besides, according to \eqref{intResult}, \eqref{OF} and \eqref{HSNR}, under the condition that $\xi_k<1$ and $N_k \leq r_k$, the diversity gain of the $k$-th user, denoted by $d_{g,k}$, satisfies that
\begin{align}
d_{g,k} & = -\lim_{\rho_{k,\min}\rightarrow+\infty}\frac{\min_{\rho_{\pmb{s}_k,\hat{\pmb{s}}_k}} \ln \mathcal{P}(\bar{\pmb Z}_k\rightarrow\hat{\bar{\pmb Z}}_k|\hat{\bar{\pmb v}}_k^H) }{\ln \rho_{k,\min}} \nonumber\\
& \geq -\lim_{\rho_{k,\min}\rightarrow+\infty}\frac{\ln \tilde{g}_k(\pmb M_k) }{\ln \rho_{k,\min}} \nonumber \\
& =\lim_{\rho_{k,\min}\rightarrow+\infty} \frac{\ln\det\left(\rho_{k,\min}\pmb M_k^H\pmb{A}_k\pmb M_k+\pmb{I}_{N_k}\right)}{\ln \rho_{k,\min}}\nonumber\\
&=\frac{N_k \ln \rho_{k,\min}+\ln\det\left(\pmb M_k^H\pmb{A}_k\pmb M_k\right)}{\ln \rho_{k,\min}}=N_k, \label{DiversityGainDerivation}
\end{align}
which coincides with the diversity gain that can be achieved by the $(N_k\times T_k)$-dimensional orthogonal space-time block code when the receiver has one antenna.

\subsubsection{Low SNR}
Under the low SNR scenario, $\sigma^2$ is large, and we have $\rho_{k,\min}\rightarrow0$. Using the Taylor expansion
$\ln\det(\pmb I+\pmb X) = \mathrm{tr}(\pmb X)+o(\left\|\pmb X\right\|)$
where $o(\cdot)$ denotes the higher order infinitesimal, and Taylor series $\left(\pmb I-\pmb X\right)^{-1} = \sum_{k=0}^{\infty}\pmb X^k$, the objective function of (\ref{optProblem}) at the low SNR regime can be approximated by
\begin{align}\label{LSNRApproximation}
\tilde{f}_k\left(\pmb M_k\right)
&\approx  - \rho_{k,\min}\pmb\omega_k^H\pmb{A}_k^{\frac{1}{2}} \pmb {M}_k\pmb M_k^H\pmb{A}_k^{\frac{1}{2}}\pmb\omega_k  + \pmb\omega_k^H\pmb\omega_k -\tilde{b}_k \nonumber \\
&\quad -\rho_{k,\min}\mathrm{tr}\left(\pmb{A}_k^{\frac{1}{2}}\pmb{M}_k\pmb M_k^H\pmb{A}_k^{\frac{1}{2}}\right).
\end{align}
After ignoring the constant term in \eqref{LSNRApproximation}, problem \eqref{optProblem} can be approximated as
\begin{align}\label{LSNR}
\max_{\left\|\pmb M_k\right\|_F^2\leq\alpha_k}\  \tilde{f}_k^{(\mathrm{L})}\left(\pmb M_k\right),
\end{align}
where $\tilde{f}_k^{(\mathrm{L})}\left(\pmb M_k\right) \triangleq  \mathrm{tr}(\pmb{A}_k^{\frac{1}{2}}\pmb{M}_k\pmb M_k^H\pmb{A}_k^{\frac{1}{2}}) + \pmb\omega_k^H\pmb{A}_k^{\frac{1}{2}} \pmb {M}_k\pmb M_k^H\pmb{A}_k^{\frac{1}{2}}\pmb\omega_k$
The optimal solution to problem \eqref{LSNR} is provided in the following proposition.
{
\begin{prop}
Assume that $\pmb \theta_{k,\max}$ is the normalized eigenvector corresponding to the maximum
eigenvalue of $\pmb\Theta_k \triangleq \pmb\Lambda_k+\pmb\Lambda_k^{1/2}\pmb \alpha_k\pmb \alpha_k^H\pmb\Lambda_k^{1/2}$, then
the optimal solution to problem \eqref{LSNR}, denoted by $\pmb M_k^o$, can be written as $\mathrm{vec}(\pmb M_k^o)=\sqrt{\alpha_k}\left(\pmb{a}\otimes \pmb{\theta}_{k,\max}\right)$, where
$\pmb{a}$ can be any $N_k$-dimensional vector,
\end{prop}
}
\begin{IEEEproof}
Due to the fact that $\mathrm{tr}\left(\pmb{X}^H\pmb{Y}\right)=\mathrm{vec}(\pmb{X})^H\mathrm{vec}(\pmb{Y})$ and $\mathrm{vec}\left(\pmb{XYZ}\right)=\left(\pmb{Z}^T\otimes\pmb{X}\right)\mathrm{vec}(\pmb{Y})$, $\tilde{f}_k^{(\mathrm{L})}\left(\pmb M_k\right)$ in \eqref{LSNR} can be reformulated as
\begin{align}
\tilde{f}_k^{(\mathrm{L})}\left(\pmb M_k\right) &=  \mathrm{tr}\left(\pmb M_k^H\left(\pmb{A} + \pmb{A}_k^{\frac{1}{2}}\pmb\omega_k \pmb\omega_k^H\pmb{A}_k^{\frac{1}{2}}\right) \pmb{M}_k\right) \nonumber \\
& =
\mathrm{vec}\left(\pmb{M}_k\right)^H
\left(\pmb{I}_{N_k}\otimes \pmb\Theta_k\right)
\mathrm{vec}\left(\pmb{M}_k\right), \label{ReformedLSNR}
\end{align}
Note that \eqref{ReformedLSNR} is in a  positive semidefinite quadratic form, and it is straight that $\mathrm{vec}(\pmb{M}_k^o) = \sqrt{\alpha_k}\pmb{u}_{k}$, where $\pmb{u}_{k}$ is the normalized eigenvector corresponding to the largest eigenvalue of matrix $\left(\pmb{I}_{N_k}\otimes \pmb\Theta_k\right)$.
Let $\zeta_{k,\max}$ denote the largest eigenvalue of $\pmb\Theta_k$, and the corresponding normalized eigenvector is $\pmb{\theta}_{k,\max}$. Then $\pmb{u}_{k}$ is in the form of $\pmb{u}_{k} = \pmb{a}\otimes \pmb{\theta}_{k,\max}$, where $\pmb{a}$ can be any normalized $N_k$-dimensional vector, because any normalized $N_k$-dimensional vector is an eigenvector of $\pmb{I}_{N_k}$.
\end{IEEEproof}
The result shows that at the low SNR regime, the optimial transmission scheme is doing beamforming at each symbol interval of the OSTBC.

\subsubsection{Without CSI}\label{withoutCSISubsection}
We now consider the scenario that the BS does not have any instantaneous CSI about $\bar{\pmb v}_k$, i.e., we have $\xi_k=0$ in (\ref{channelEstModel}). Then, by ignoring the constant term, \eqref{optProblem} becomes
\begin{align}\label{withoutCSI}
\max_{\left\|\pmb M_k\right\|_F^2\leq\alpha_k}&\quad \ln\det\left(\rho_{k,\min}\pmb M_k^H\pmb\Lambda_k\pmb M_k+\pmb{I}_{r_k}\right).
\end{align}
Note that optimization problem \eqref{withoutCSI} has  the same mathematical structure as that in \eqref{infSNR}, and therefore, the optimal solution of \eqref{withoutCSI} follows the water-filling principle in \eqref{waterfillingsolution}, i.e., the optimal scheme is to allocate power to OSTBC streams according to the water-filling principle according to the statistical CSI $\pmb\Lambda_k$.

\section{Sum Power Constraint}\label{sumSection}
Section \ref{indSection} considers the individual user power constraint where the linear weight matrix $\pmb{W}$ is divided into $K$ sub-matrices which are designed independently to minimize the PEP of each user under the corresponding power constraint. In this section, we discuss the sum power constraint of all $K$ users, and design $\pmb{W}_k, k=1,2,\cdots,K$ jointly.

Since the performance of all the $K$ users are optimized together, we consider two performance criterions, i.e., the $\min-\max$ PEP and the average PEP problems. Specially, our target is to minimize the largest PEP and the average PEP of $K$ users under the sum power constraint respectively. We still focus on optimizing the upper bound of PEP in (\ref{intResult}) in the following.

\subsection{$\min-\max$ PEP Problem}
Under this criterion, we design $\pmb M_k, k=1,2,\cdots,K$ to minimize the worst-user PEP.  Using the upper bound (\ref{intResult}), the  optimization problem is written as
\begin{align}
\min_{\mathcal{M}\in \mathcal{D}}&\quad\max_{k=1,\cdots,K}\tilde{g}_k\left(\pmb M_k\right),
\end{align}
where
$\tilde{g}_k\left(\pmb M_k\right)$ is defined in \eqref{OF}, $\mathcal{M}\triangleq\left(\pmb{M}_1,\pmb{M}_2,\cdots,\pmb{M}_K\right)$, and
$\mathcal{D}\triangleq\Big\{\mathcal{M} |\sum_{k=1}^K\frac{L_k}{T}\left\|\pmb{M}_k\right\|_F^2\leq P_{T,\max}\Big\}$.
By Introducing a slack variable $t$ and taking the logarithm with respect to $\tilde{g}_k(\pmb M_k)$ for $1\leq k\leq K$, the problem is equivalent to
\begin{align}
\label{minMaxFirst}
\min_{ t,\mathcal{M}\in \mathcal{D}}&\  t, \quad \mathrm{s.t.}\ \tilde f_k(\pmb M_k)\leq\ln( t),~ \forall k=1,2,\cdots,K,
\end{align}
The problem \eqref{minMaxFirst} is not a convex problem due to the non-convex constraints. Before solving the problem, we first esstablish the average PEP problem.

\subsection{Average PEP Problem}
For this performance criterion, our target is to minimize the average PEP of $K$ users. The optimization problem can be written as
\begin{align}
\label{OriginalPEP}
\min_{\mathcal{M}\in \mathcal{D}}  \quad \frac{1}{K}\sum_{k=1}^K\tilde{g}_k(\pmb M_k).
\end{align}
Introducing  $t_k, k=1,2,\cdots,K$, we can obtain an equivalent optimization problem as follow,
\begin{align}
\label{AOOrignalFirst}
\min_{t_k,\mathcal{M}\in \mathcal{D}}\ \sum_{k=1}^Kt_k,\mathrm{s.t.}\ \tilde{f}_k(\pmb M_k)\leq\ln( t_k),\forall k=1,2,\cdots,K.
\end{align}

We can see that both the problems \eqref{minMaxFirst} and \eqref{AOOrignalFirst} have the similar forms. The difficulty of solving them lies in the fact that $\tilde{f}_k(\pmb M_k)$ for $1\leq k\leq K$ are non-convex functions. Similar to the previous discussions in the individual user power constraint case, SDR and Gaussian Randomization method could be exploited to solve them by taking $\pmb\Omega_k=\pmb M_k\pmb M_k^H$ and impose the rank constraints $\mathrm{rank}(\pmb\Omega_k)\leq N_k, k=1,\cdots,K$.

However, compared to the individual case, the problem dimension is much larger. Besides, if the optimal solutions obtained by SDR do not satisfy the rank constraints, Gaussian Randomization should be used among $K$ optimization variables, which is more difficult to be satisfied. Hence, in the following, we provide a more efficient method, which can be used to solve both of \eqref{minMaxFirst} and \eqref{AOOrignalFirst}. The basic idea of our method is that we first transform the complicated non-convex optimization problem into a sequence of convex problems by SCA method. Further more, we using the ADMM algorithm to divide the obtained convex
problem into several subproblems which can be implemented in a parallel manner, and thus significantly reduces the complexity of solving the problem.

\subsection{SCA-ADMM Method}
In this subsection, we solve the optimization problems established in previous two subsections. For brevity, we take the average PEP problem as an example and the $\min-\max$ PEP problem can be solved in a similar manner.
Analogue to the individual user power constraint, for the sum power constraint, the optimization problem can still be solved by SCA method. In the $i$-th iteration, we have the following optimization problem,
\begin{align}
\label{SCAProbSumPower}
\begin{aligned}
\min_{ t_k,\mathcal{M}\in \mathcal{D}}&\quad \sum_{k=1}^Kt_k,\\
\mathrm{s.t.}\ \ &\quad \hat{f}_{k}^{(i)}\left(\pmb M_k\right) +c_k\leq\ln( t_k),~\forall k = 1,2,\cdots,K,
\end{aligned}
\end{align}
where $c_k$ and $\hat{f}_{k,j}\left(\pmb M_k,\hat{\pmb M}_{k,i}\right)$ for $j\in\{1,2\}$ are defined in \eqref{SCAIgnoreConstant} and  \eqref{firstUpperBound}, respectively.
Note that \eqref{SCAProbSumPower} is a convex problem.
In the following, we solve \eqref{SCAProbSumPower} by using the ADMM algorithm \cite{S.Boyd2011}. The main advantage of the ADMM algorithm is that it decomposes  \eqref{SCAProbSumPower} into several subproblems which can be implemented in a parallel manner, and therefore significantly reduces the time consumption if the BS is equipped with multiple computing units.

First, we rewrite \eqref{SCAProbSumPower} as \eqref{SCAADMM}, which is given at the top of the next page,
\begin{figure*}[t]
\begin{subequations}
\label{SCAADMM}
\begin{align}
\min_{ t_k,\tau_k,\mathcal{M}\in \mathcal{D},\mathcal{X}}&\quad \sum_{k=1}^K\left(
\tau_k
+\frac{\delta}{2}\left\|\pmb{M}_k - \pmb{X}_k + \pmb{V}_k\right\|_F^2 +\frac{\delta}{2}\left(t_k - \tau_k + \chi_k\right)^2\right),\\
\mathrm{s.t.}\quad \ &\quad \hat{f}_{k}^{(i)}\left(\pmb M_k\right) +c_k\leq\ln( \tau_k),\quad\forall k = 1,2,\cdots,K,\\
&\quad \pmb{M}_k = \pmb{X}_k,\ t_k = \tau_k, \quad \forall k = 1,2,\cdots,K,
\end{align}
\end{subequations}
\rule{\textwidth}{.2mm}
\vspace{-7mm}
\end{figure*}
where $\mathcal{X} \triangleq \left(\pmb{X}_1,\pmb{X}_2,\cdots,\pmb{X}_K\right)$, $\delta>0$ is a constant, $\pmb{V}_k$ for $k=1,2,\cdots,K$ are any compatible matrix, and $\chi_k$ for $k=1,2,\cdots,K$ are some real numbers.

Using the principle of ADMM, \eqref{SCAADMM} can be solved by iterating the following steps:
\begin{enumerate}
\item For $k=1,2,\cdots,K$, update $\{\pmb{X}_k,\tau_k\}$ with $\{\pmb{M}_k,t_k,\pmb{V}_k,\chi_k\}_{k=1}^K$ fixed as constants by solving the following optimization problem,
\begin{align}
\label{ADMMStep1}
\begin{aligned}
\min_{\tau_k, \pmb{X}_k}\ & \tau_k + \frac{\delta}{2}\left\|\pmb{M}_k - \pmb{X}_k + \pmb{V}_k\right\|_F^2+\frac{\delta}{2}\left(t_k - \tau_k + \chi_k\right)^2,\\
\mathrm{s.t.}\  &\hat{f}_{k}^{(i)}\left(\pmb M_k\right) + c_k\leq\ln( \tau_k ),
\end{aligned}
\end{align}
which is a convex optimization problem. Besides, we can solve $\{\pmb{X}_k,\tau_k\}$ for different $k$ in a parallel manner. We show that this optimization problem can be efficiently solved by a simple bisection method in the Appendix D.

\item For $k=1,2,\cdots,K$, update $\{\pmb{M}_k,t_k\}$  with $\{\pmb{X}_k,\tau_k,\pmb{V}_k,\chi_k\}_{k=1}^K$ fixed as constants by solving the following problem,
\begin{align}
\min_{t_k,\mathcal{M}\in\mathcal{D} }\  \sum_{k=1}^K\Big(
\left\|\pmb{M}_k-\pmb{X}_k+\pmb{V}_k\right\|_F^2 +\left\|t_k - \tau_k + \chi_k\right\|^2\Big).\nonumber
\end{align}
Note that this is a QCQP-1 problem which can be solved by bisection method \cite{K.Huang2016}. For a special case, where $L_1=L_2=\cdots=L_K=L$,
denote the optimal solution as $\{t_{k,\mathrm{opt}},\pmb{M}_{k,\mathrm{opt}}\}_{k=1}^K$, then we have
$t_{k,\mathrm{opt}} = \tau_k+\chi_k$, and $\bar{\pmb{m}}_{\mathrm{opt}}=\bar{\pmb{x}}-\bar{\pmb{v}}$ if $\left\|\bar{\pmb{x}}-\bar{\pmb{v}}\right\|^2\leq \frac{TP_{T,\max}}{L}$, otherwise, $\bar{\pmb{m}}_{\mathrm{opt}} = \frac{\bar{\pmb{x}}-\bar{\pmb{v}}}{\left\|\bar{\pmb{x}}-\bar{\pmb{v}}\right\|}$,
where
$\bar{\pmb{m}}_{\mathrm{opt}}
\triangleq[\mathrm{vec}\left(\pmb{M}_{1,\mathrm{opt}}\right)^H,
\cdots,\mathrm{vec}\left(\pmb{M}_{K,\mathrm{opt}}\right)^H]^H$,
$\bar{\pmb{x}}
\triangleq[\mathrm{vec}\left(\pmb{X}_{1}\right)^H, \cdots,\mathrm{vec}\left(\pmb{X}_{K}\right)^H]^H$, and
$\bar{\pmb{v}}
\triangleq[\mathrm{vec}\left(\pmb{V}_{1}\right)^H, \cdots,\mathrm{vec}\left(\pmb{V}_{K}\right)^H]^H$.

\item Update $\{\pmb{V}_k,\chi_k\}$ for $k=1,2,\cdots,K$ as follows,
\begin{align}
\pmb{V}_k\leftarrow \pmb{V}_k + \left(\pmb{M}_k-\pmb{X}_k\right),
\chi_k\leftarrow \chi_k + \left(t_k-\tau_k\right).
\end{align}
\end{enumerate}
As shown in \cite{S.Boyd2011}, $\{\pmb{M}_k,t_k\}$ in the above iteration will finally converge to the global optimal point of \eqref{SCAADMM}, and thus the global optimal point of \eqref{SCAProbSumPower}.
By iteratively solving \eqref{SCAProbSumPower}, we can obtain a local optimal point of \eqref{AOOrignalFirst}, i.e., solving the average PEP problem. The $\min-\max$ PEP problem can be solved in a similar manner. We only need to replace the constraint $\tau_k = t_k$ in \eqref{SCAADMM} with $\tau_k = t$, and the subsequent steps remain nearly the same.

\section{Simulation Results}\label{simSection}
\begin{figure}[t]
\begin{center}
\includegraphics[width=2.7 in]{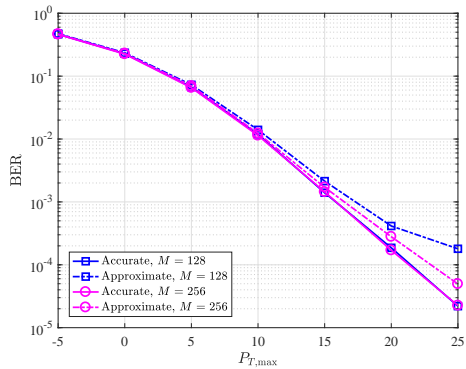}
\end{center}
\caption{\small BERs of the DFT approximation method in \eqref{proposition} and the actual eigenvalue decomposition, where we set $\Delta=10^\circ$ and $K=4$.}
\label{compasionDFTEVD}
\vspace{-5mm}
\end{figure}

In this section, some simulation results are provided to evaluate the performance of our proposed method. Unless specified, for each $\pmb Z_k$, we consider the well-known Alamouti space-time code \cite{Alamouti} which means that  $N_k=L_k=T=2$ for $k=1,2,\cdots,K$. All signal symbols are assumed to be the QPSK symbol. We set $\Delta=\Delta_1=\Delta_2=\cdots=\Delta_K$ and $\xi=\xi_1=\xi_2=\cdots=\xi_K$ for illustrative convenience. The mean azimuth angles of the users are set to be evenly spaced within $[-60^\circ,60^\circ]$. When individual user power constraint is considered, we set $\alpha_k=P_{T,\max}/K, k=1,2,\cdots,K$, i.e., the power are uniformly allocated among different users.

\subsection{BER when the BS has no CSI}
In Fig. \ref{compasionDFTEVD}, we plot the BERs when the BS has no prior knowledge of the instantaneous CSI. We compare the BERs between the DFT approximation in \eqref{proposition} and the actual eigenvalue decomposition under the individual user power constraint. As we can see from Fig. \ref{compasionDFTEVD}, as $M\rightarrow\infty$, the BERs of the DFT approximation method is approaching to the actual performance. It suggests that using \eqref{proposition} to obtain the eigenvalues and eigenvectors of the channel covariance matrices will reduce the computational complexity while maintaining almost the same BER performance, when $M$ is large.
Besides, we also see that under the condition that the BS has no prior knowledge of the CSI, the increase of $M$ has little impact on improving the BER. Actually, if \eqref{character} is satisfied, $M$ plays almost no effect on the designs of $\pmb M_k$ as shown in Section \ref{withoutCSISubsection}, which indicates that using only the second-order statistics of the channel is generally ineffective.

\begin{figure}[t]
  \centering
  \subfigure[The convergence of the SCA method.]{
    \label{SCAConvergence} 
    \includegraphics[width=2.7 in]{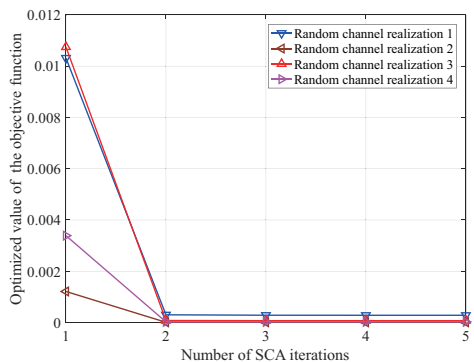}}
  \subfigure[The convergence of the ADMM algorithm.]{
    \label{ADMMConvergence} 
    \includegraphics[width=2.7 in]{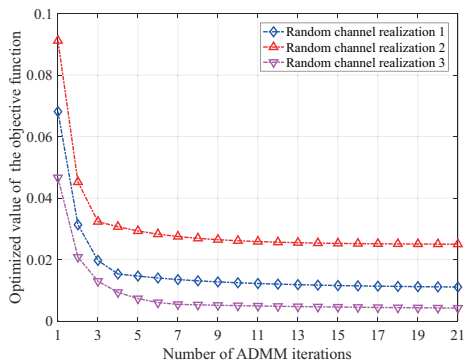}}
  \caption{The convergence of the proposed method.}
  \label{WholeConvergence} 
  \vspace{-5mm}
\end{figure}
\subsection{BER when the BS has partial CSIs}

In this part, we check the system performance in terms of the BER when partial CSIs are obtained at the BS.

In Fig. \ref{WholeConvergence}, we evaluate the convergence performance of the SCA and the ADMM algorithm. We set $M = 128$ and $K=2$ in the simulation.
In Fig. \ref{SCAConvergence}, we plot the optimized values of the objective function versus the SCA iteration steps when solving solving the average PEP problem in \eqref{OriginalPEP}. As we can see, the SCA method converges within a few number of iterations.
In Fig. \ref{ADMMConvergence}, the convergence of the ADMM algorithm when solving \eqref{SCAProbSumPower} is plotted. In fact, in the most of our simulations, the ADMM algorithm usually converges within tens of iterations. Note that when using the ADMM algorithm, the updates of $\{t_k,\pmb{M}_k\}_{k=1}^K$ and $\{\pmb{V}_k,\chi_k\}_{k=1}^K$ are obtained in closed form. Besides, the updates of $\{\tau_k,\pmb{X}_k\}_{k=1}^K$ can be solved by a simple bisection method in a parallel manner. Therefore, using the ADMM algorithm to solve  \eqref{SCAProbSumPower} is computationally efficient.

\begin{figure}[t]
\centering
\includegraphics[width=2.7 in]{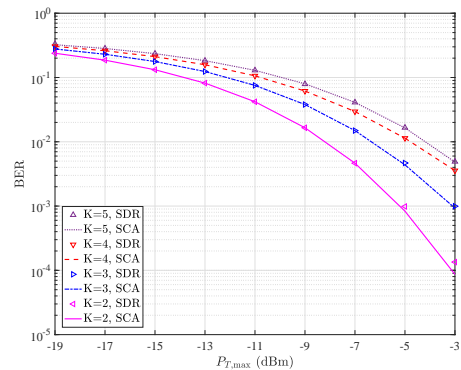}
\caption{BER versus $P_{T,\max}$ under individual user power constraint, where $M = 128$, $\xi = 0.8$, and $\Delta=7.5^\circ$.}\label{IndividualPowerForDifferentUsers}
\vspace{-5mm}
\end{figure}

In Fig. \ref{IndividualPowerForDifferentUsers}, we plot the BER under the individual user power constraint.
In the simulation, we use two different methods to obtain the beamforming matrices, i.e., the SDR method and the SCA method proposed in Section \ref{SDRIndividualPower} and Section \ref{SCAIndividualPower}, respectively.
In the SDR method, if the optimization results do not meet the rank constraint, then we use the Gaussian randomization technique to recover a proper solution (selected from 1000 randomly generated samples). As we can see, the two methods achieve similar performance. However, in our simulation settings, the SCA method is  computationally much more efficient than the SDR method. This is because we generally have $r_k\gg2$, and the number of variables in the SDR method is $\sum_{k=1}^K r_k^2$, which is much larger than that in the SCA method, i.e., $\sum_{k=1}^K 2r_k$, and in each iteration in the SCA method, the optimization problem becomes a QCQP-1 can be efficiently solved by a bisection method.

\begin{figure}[t]
\centering
\includegraphics[width=2.7 in]{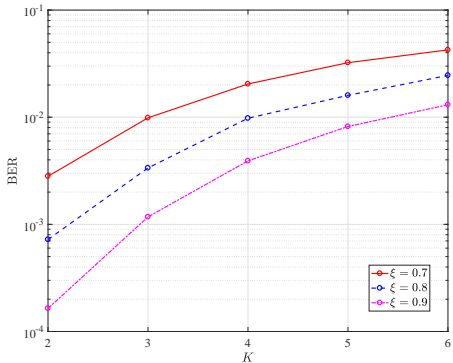}
\caption{BER versus $K$ under individual power constraint, where $M = 256$, $\Delta = 5^{\circ} $, and $P_{T,\max} = -8$ (dBm).}
\label{BER2}
\vspace{-5mm}
\end{figure}

\begin{figure}[t]
\centering
\includegraphics[width=2.7 in]{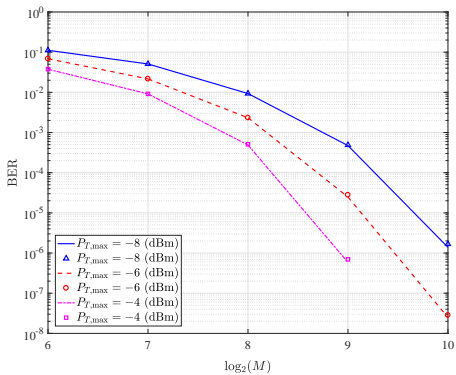}
\caption{BER versus the number of antennas under individual power constraint, where $\xi = 0.8$, $K = 5$, and $\Delta = 5^{\circ}$.}
\label{BER3}
\vspace{-5mm}
\end{figure}

Fig. \ref{BER2}, we plot the BERs versus the numbers of the users, where we set $\xi$ as $0.7$, $0.8$, and $0.9$. As shown in Fig. \ref{BER2}, with the increase of $K$, the BER also increase due to the fact that the power for each user is decreased. Besides, due to the fact that the physical channels are not perfectly orthogonal to each other, the interfering power among adjacent users also increase with $K$, which also results in the increase of the BER.

Fig. \ref{BER3} illustrates the BERs under different numbers of antennas at the BS, i.e., $M$.
As we can see,  the increase of $M$ results in a significant decrease of the BER. Note that this is different from the results in Fig. \ref{compasionDFTEVD} wherein the BS does not have any prior knowledge of the CSI and the BER improves little by increasing $M$. Here, with partial CSIs, the BS is able to align (even though imperfectly) its beamforming matrices with the channels, and brings the antennas power gains at the users. Therefore, with the partial CSIs, the BER decreases with the increase of $M$.

\begin{figure}[t]
\centering
\includegraphics[width=2.7 in]{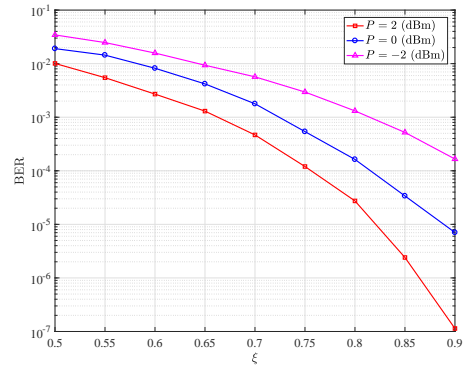}
\caption{\small BER versus the channel correlation coefficient $\xi$ under sum power constraint, $M=128$, $K=4$ and $\Delta = 7.5^{\circ}$.}
\label{BERCurve2Xi}
\vspace{-5mm}
\end{figure}

Fig. \ref{BERCurve2Xi} illustrates the BERs against the channel correlation coefficients, i.e., $\xi$.
In the simulation, we consider the sum power constraint, and the power allocation among different user are obtained by solving \eqref{OriginalPEP}. According to \eqref{channelEstModel}, a larger value of $\xi$ means that the CSIs at the BS is more accurate. From the simulations results in Fig. \ref{BERCurve2Xi}, we can see that the BERs can be greatly reduced with the increase of $\xi$, especially when the power budget of the BS is large.

\begin{figure}[t]
\centering
\includegraphics[width=2.7 in]{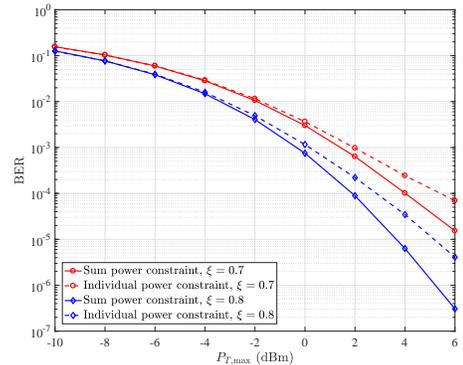}
\caption{\small BER comparison between individual power constraint and  sum power constraint, $M = 128$, $K=6$, and $\Delta = 7.5^{\circ}$.}
\label{BER1}
\vspace{-5mm}
\end{figure}

In Fig. \ref{BER1}, we compare the BERs when the BS uniformly allocates it power to different users and when the BS jointly designs the beamforming matrices and the power allocation. As we can see from Fig. \ref{BER1}, the BER performance under the sum power constraint outperforms that under the individual power constraint because the power allocation are optimized. { However, we also see that if $P_{T,\max}$ is small, then the two constraints lead to similar BER performance. This inspires us that when the transmit power of the BS is low, allocating the power uniformly among different users  is a good method to simplify the design of the beamforming matrices while ensuring little performance loss.}

\subsection{Comparison with JSDM scheme}
\begin{figure}[t]
\centering
\includegraphics[width=2.7 in]{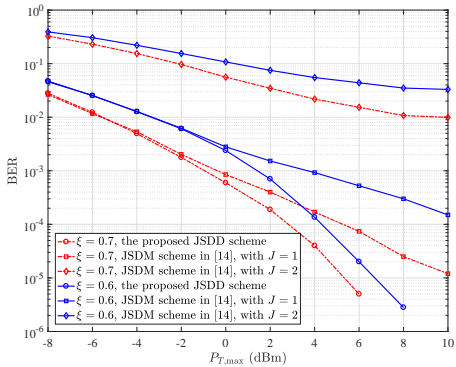}
\caption{BER comparison between JSDD and JSDM, where we adopt the $8\times 8$ OSTBC for the JSDD scheme.}\label{JSDMJSDD}
\vspace{-5mm}
\end{figure}

\begin{figure}[t]
\centering
\includegraphics[width=2.7 in]{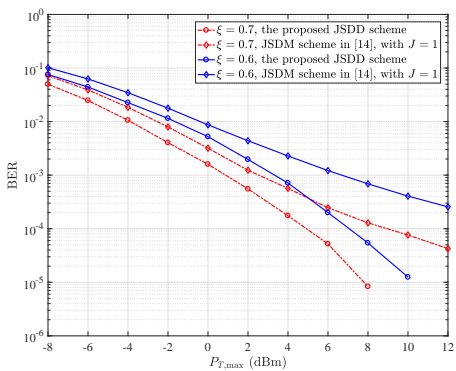}
\caption{BER comparison between JSDD and JSDM, where we adopt the $4\times 4$ OSTBC for the JSDD scheme.}\label{JSDMJSDD2}
\vspace{-5mm}
\end{figure}

In this subsection, we compare the proposed JSDD scheme with the JSDM scheme proposed in \cite{covMatForm} in term of the BER when the BS only has partial CSIs. The simulation results are illustrated in Fig. \ref{JSDMJSDD} and Fig. \ref{JSDMJSDD2}. In the simulation, we set $M=128$, $K=4$, $\Delta = 5^{\circ}$. Note that the JSDM scheme can serve more than $1$ user in each user group, i.e., $J\geq 1$, and therefore, we include both the case where $J=1$ and where $J=2$ in our simulation, which are referred to as JSDM-1 and JSDM-2, respectively. The signal constellations are the BPSK and QPSK in Fig. \ref{JSDMJSDD} and Fig. \ref{JSDMJSDD2}, respectively.
For the proposed JSDD scheme, we adopt the $8\times 8$ and the $4\times 4$ OSTBC in Fig. \ref{JSDMJSDD} and in Fig. \ref{JSDMJSDD2}, respectively \footnote{  For the construction of the mentioned OSTBC matrix, please see \cite{SpaceTimeOrthgonal}.}.
For computational efficiency in our simulation, we adopt the individual user power constraint for the proposed JSDD scheme.

As we can see in Fig. \ref{JSDMJSDD}, for the case of JSDM-2, the BERs are generally very high, and even the BS adopts a high transmit power, the BER seems to be bounded by the decoding error floor.
This is because the CSIs at the BS are imperfect, and the decoding error floor is caused by the intra-group interference.
For the case of JSDM-1, there is no intra-group interference, and therefore, the BER keep decreasing with the increase of the transmit power. Note that this means that under the condition of imperfect CSIT, the JSDM scheme may be more suitable to serve only one user in each group.
Both Fig. \ref{JSDMJSDD} and Fig. \ref{JSDMJSDD2} reveal the fact when the CSIT is inaccurate, i.e., $\xi = 0.6~\text{and}~0.7$, the proposed JSDD scheme outperforms the JSDM-1 scheme in term of BER. This is mainly because with the utilization of the OSTBC, the JSDD scheme can achieve diversity gains,  but the beams in the JSDM-1 scheme are formed towards the wrong directions and the effective received power at the users is significantly reduced.

\section{Conclusion}\label{conclusionSection}
In this paper, we propose a JSDD downlink transmission for multi-user massive MIMO system. The basic idea is to permit $K$ users to access the downlink transmissions via spatial division by utilizing the low-rank property of channels, and for each user OSTBC with partial CSI (estimated CSI with errors) is utilized to provide diversity gain. We provide detailed design under both individual user power and sum power constraint,  SDR method and SCA-ADMM algorithm are proposed to solve the optimization problem. Some special cases with closed-form solutions are also discussed. The JSDD scheme is robust to CSIT errors and provides high reliability for downlink transmission. The scheme could find its applications in URLLC scenarios, multiple-group multicast scenarios, etc.

\appendix

\subsection{Proof of Proposition 2}
The corresponding KKT conditions on $\pmb\Omega_k$ are
\begin{align}
\frac{\partial L}{\partial{\pmb\Omega_k}}
&=\ -\rho_{k,\min}\tilde{\pmb B}_k^T(\pmb\Omega_k)-\pmb Q+\kappa_2\pmb I \nonumber \\
&\quad -\kappa_1\rho_{k,\min}\left[\tilde{\pmb B}_k(\pmb\Omega_k)\pmb\mu_k\pmb\mu_k^H\tilde{\pmb B}_k(\pmb\Omega_k)\right]^{T}=\pmb 0, \label{KKTCondFirst}\\
\pmb Q\pmb\Omega_k &= \pmb 0,\label{KKTCond}
\end{align}
From \eqref{KKTCond}, we have
$
0=\mathrm{rank}(\pmb Q\pmb\Omega_k)\geq\mathrm{rank}(\pmb Q)+\mathrm{rank}(\pmb \Omega_k)-r_k
$,
i.e.,
$\mathrm{rank}(\pmb \Omega_k)\leq r_k-\mathrm{rank}(\pmb Q)$.
Furthermore, \eqref{KKTCondFirst} shows that
$\pmb Q = \kappa_2\pmb I-\rho_{k,\min}\tilde{\pmb B}_k^{T}(\pmb\Omega_k)-\kappa_1\rho_{k,\min}\left[\tilde{\pmb B}_k(\pmb\Omega_k)\pmb\mu_k\pmb\mu_k^H\tilde{\pmb B}_k(\pmb\Omega_k)\right]^{T}$.
Since $\mathrm{rank}(\kappa_1\rho_{k,\min}[\tilde{\pmb B}_k(\pmb\Omega_k)\pmb\mu_k\pmb\mu_k^H\tilde{\pmb B}_k(\pmb\Omega_k)]^{T})= 1$, then we have $
\mathrm{rank}(\pmb Q)\geq\mathrm{rank}\left(\kappa_2\pmb I-\rho_{k,\min}\tilde{\pmb B}_k^{T}(\pmb\Omega_k)\right)-1$,
which leads to the conclusion in proposition 2.

\subsection{Derivation of \eqref{firstUpperBound}}
Before deriving \eqref{firstUpperBound}, we first provide an useful inequality:
for any two matrices $\pmb {X}$ and $\pmb{Y}$ with compatible dimensions and positive definite $\pmb Y$, the following inequality holds
\begin{align}
&\mathrm{tr}\left(\pmb{XY}^{-1}\pmb X^H\right)\nonumber \\
&\geq\mathrm{tr}\left(\hat{\pmb X}\hat{\pmb Y}^{-1}\hat{\pmb X}^H\right)+2\mathrm{Re}\left(\mathrm{tr}\left(\left(\pmb X-\hat{\pmb X}\right)\hat{\pmb Y}^{-1}\hat{\pmb X}^H\right)\right)
\nonumber \\
&\quad -\mathrm{tr}\left(\hat{\pmb X}\hat{\pmb Y}^{-1}\left({\pmb Y}-\hat{\pmb Y}\right)\hat{\pmb Y}^{-1}\hat{\pmb X}^H\right), \label{BaseInequality}
\end{align}
where $\hat{\pmb X}, \hat{\pmb Y}$ are two constant matrices and $\hat{\pmb Y}$ is positive definite.
The proof  simply follows the fact that $\mathrm{tr}\left(\pmb{XY}^{-1}\pmb X^H\right)$ is jointly convex w.r.t. $(\pmb{X},\pmb{Y})$, and Jensen's inequality.
Based on \eqref{BaseInequality}, for $\tilde{f}_{k,1}\left(\pmb M_k\right)$, we have \eqref{SCAFucntion1}
\begin{figure*}[t]
\begin{align}
\label{SCAFucntion1}
\tilde{f}_{k,1}\left(\pmb M_k\right)&=-\pmb\omega_k^H\pmb C_k\pmb{M}_k\pmb G_k^{-1}\pmb M_k^H\pmb C_k^H\pmb\omega_k\nonumber\\
&\leq -\pmb\omega_k^H\pmb C_k\left(\pmb{M}_k-\hat{\pmb{M}}_{k,i}\right)\hat{\pmb G}_{k,i}^{-1}\hat{\pmb M}_{k,i}^H\pmb C_k^H\pmb\omega_k -\pmb\omega_k^H\pmb C_k\hat{\pmb{M}}_{k,i}\hat{\pmb G}_{k,i}^{-1}\left(\pmb M_k^H-\hat{\pmb M}_{k,i}^H\right)\pmb C_k^H\pmb\omega_k
\nonumber \\
& \quad +\pmb\omega_k^H\pmb C_k\hat{\pmb{M}}_{k,i}\hat{\pmb G}_{k,i}^{-1}\left(\pmb G_k-\hat{\pmb G}_{k,i}\right)\hat{\pmb G}_{k,i}^{-1}\hat{\pmb M}_{k,i}^H\pmb C_k^H\pmb\omega_k
-\pmb\omega_k^H\pmb C_k\hat{\pmb{M}}_{k,i}\hat{\pmb G}_{k,i}^{-1}\hat{\pmb M}_{k,i}^H\pmb C_k^H\pmb\omega_k\nonumber\\
&=-2\mathrm{Re}\left(\pmb\omega_k^H\pmb C_k\pmb{M}_k\hat{\pmb G}_{k,i}^{-1}\hat{\pmb M}_{k,i}^H\pmb C_k^H\pmb\omega_k\right)
+\pmb\omega_k^H\pmb C_k\hat{\pmb{M}}_{k,i}\hat{\pmb G}_{k,i}^{-1}\pmb G_k\hat{\pmb G}_{k,i}^{-1}\hat{\pmb M}_{k,i}^H\pmb C_k^H\pmb\omega_k =\hat{f}_{k,1}\left(\pmb M_k,\hat{\pmb M}_{k,i}\right),
\end{align}
\rule{\textwidth}{.2mm}
\vspace{-7mm}
\end{figure*}
where $\hat{\pmb M}_{k,i}$ is a constant matrix in the $i$-th iteration of the SCA method, $\hat{\pmb G}_{k,i}=\hat{\pmb M}_{k,i}^H\pmb C_k^H\pmb C_k\hat{\pmb{M}}_{k,i}+\pmb I$.
Similarly, for $\tilde{f}_{k,2}\left(\pmb M_k\right)$, we have \eqref{SCAFucntion2}
\begin{figure*}[t]
\begin{align}
\label{SCAFucntion2}
&\tilde{f}_{k,2}\left(\pmb M_k\right)=\ln\det(\pmb P_k) \overset{(a)}{\leq}\ln\det(\hat{\pmb P}_{k,i})+\mathrm{tr}\left(\hat{\pmb P}_{k,i}^{-1}\left(\pmb P_k-\hat{\pmb P}_{k,i}\right)\right)\nonumber\\
&=\ln\det(\hat{\pmb P}_{k,i})-\mathrm{tr}(\pmb I)+\mathrm{tr}\left(\hat{\pmb P}_{k,i}^{-1}\pmb P_k\right)\nonumber \\
&\overset{(b)}{\leq}\ln\det(\hat{\pmb P}_{k,i})-\mathrm{tr}(\pmb I)+\mathrm{tr}\left(\hat{\pmb P}_{k,i}^{-1}\right)
-\mathrm{tr}\left(\hat{\pmb P}_{k,i}^{-1}\pmb C_k\left(\pmb{M}_k-\hat{\pmb{M}}_{k,i}\right)\hat{\pmb G}_{k,i}^{-1}\hat{\pmb M}_{k,i}^H\pmb C_k^H\right)\nonumber\\
&\quad-\mathrm{tr}\left(\hat{\pmb P}_{k,i}^{-1}\pmb C_k\hat{\pmb{M}}_{k,i}\hat{\pmb G}_{k,i}^{-1}\left(\pmb M_k^H-\hat{\pmb M}_{k,i}^H\right)\pmb C_k^H\right)
+\mathrm{tr}\left(\hat{\pmb P}_{k,i}^{-1}\pmb C_k\hat{\pmb{M}}_{k,i}\hat{\pmb G}_{k,i}^{-1}\left(\pmb G_k-\hat{\pmb G}_{k,i}\right)\hat{\pmb G}_{k,i}^{-1}\hat{\pmb M}_{k,i}^H\pmb C_k^H\right)\nonumber\\
&\quad-\mathrm{tr}\left(\hat{\pmb P}_{k,i}^{-1}\pmb C_k\hat{\pmb{M}}_{k,i}\hat{\pmb G}_{k,i}^{-1}\hat{\pmb M}_{k,i}^H\pmb C_k^H\right)
=\hat{f}_{k,2}\left(\pmb M_k,\hat{\pmb M}_{k,i}\right),
\end{align}
\rule{\textwidth}{.2mm}
\vspace{-7mm}
\end{figure*}
where $\hat{\pmb P}_{k,i}\triangleq\pmb I-\pmb C_k\hat{\pmb{M}}_{k,i}\left(\hat{\pmb M}_{k,i}^H\pmb C_k^H\pmb C_k\hat{\pmb{M}}_{k,i}+\pmb I\right)^{-1}\hat{\pmb M}_{k,i}^H\pmb C_k^H$, step $(a)$ is because $\mathrm{log~det}(\cdot)$ is a concave function, and step $(b)$ is obtained by using \eqref{BaseInequality}.

\subsection{A bisection method to solve \eqref{ADMMStep1}}
To solve \eqref{ADMMStep1}, we first rewritten  it as the following form,
\begin{align}
\label{GeneralADMMStep1}
\begin{aligned}
\min_{\tau_k, \pmb{x}_k}&\ \ \tau_k+
\frac{\delta}{2}\left\|\pmb{x}_k-\pmb{v}_k\right\|^2
+\frac{\delta}{2}\left(\tau_k - q_k\right)^2,\\
\mathrm{s.t.}&\ \ \varphi_{k,i}\left(\pmb{x}_k\right)\leq\ln( \tau_k),
\end{aligned}
\end{align}
where
$\varphi_{k,i}\left(\pmb{x}\right)\triangleq\pmb{x}^H\pmb{\Psi}_{k,i}\pmb{x} - 2\Re\{\pmb{\beta}_{k,i}^H\pmb{x}\} + d_{k,i}$,
$\pmb{x}_k=\mathrm{vec}\left(\pmb{X}_k\right)$,
$\pmb{v}_k=\mathrm{vec}\left(\pmb{M}_k+\pmb{V}_k\right)$,
$q_k=t_k+\chi_k$, $\pmb{\Psi}_{k,i}  =
\left(\pmb{\gamma}_{k,i}^*\pmb{\gamma}_k^T + \pmb{\Xi}_{k,i}^T \right)\otimes\left( \pmb{C}_k^H\pmb{C}_k\right)$,
$\pmb{\beta}_{k,i} = \mathrm{vec}\left(\pmb{\Gamma}_{k,i}\right)$, and $d_{k,i} = c_k + a_{k,i}^{(1)} + a_{k,i}^{(2)}$.
Denote $\tilde{\tau}_{k,\mathrm{opt}}$ as the optimal solution of  \eqref{GeneralADMMStep1} when  the inequality constraint is absent, then it can be easily obtained that $\tilde{\tau}_{k,\mathrm{opt}}=\frac{\delta q_k - 1}{\delta}$.
If $\delta q_k - 1>0$ and $\varphi_{k,i}\left(\pmb{v}_k\right)\leq \ln(\tilde{\tau}_{k,\mathrm{opt}})$, then the optimal solution of \eqref{GeneralADMMStep1} is given by $\pmb{x}_{k,\mathrm{opt}} = \pmb{v}_k$ and $\tau_{k,\mathrm{opt}}=\tilde{\tau}_{k,\mathrm{opt}}$.
Otherwise, to obtain $\pmb{x}_{k,\mathrm{opt}}$ and $\tau_{k,\mathrm{opt}}$, we write the KKT conditions of \eqref{GeneralADMMStep1} as:
(a) $\pmb{x}_k\left(\lambda_k\right) = \left(\left(\delta/2\right)\pmb{I}+\lambda_k\pmb{\Psi}_{k,i}\right)^{-1}
\left(\left(\delta/2\right)\pmb{v}_k+\lambda_k\pmb{\beta}_{k,i}\right)$,
(b) $\tau_k\left(\lambda_k\right) =\left(
\delta q_k-1 + \sqrt{\left(\delta q_k-1\right)^2 + 4\delta\lambda_k}\right)\big/2\delta$,
(c) $\lambda_k \left[\varphi_{k,i}(\pmb{x}_k\left(\lambda_k\right))-\ln\left(\tau_k\left(\lambda_k\right)\right) \right] =0$, and
(d) $\varphi_{k,i}(\pmb{x}_k\left(\lambda_k\right))-\ln\left(\tau_k\left(\lambda_k\right)\right)  \leq 0, \quad \lambda_k\geq 0$,
where $\lambda_k$ is the dual variable with respect to the inequality constraint.
It can be easily checked by the first order derivative that $\phi\left(\lambda_k\right)\triangleq\varphi_{k,i}(\pmb{x}_k\left(\lambda_k\right))-\ln\left(\tau_k\left(\lambda_k\right)\right) $ is a monotonically decreasing function with respect to $\lambda_k$, and we have $\phi\left(\lambda_k\right)\rightarrow+\infty$ as $\lambda_k\rightarrow 0^{+}$. Therefore, the optimal dual variable $\lambda_{k,\mathrm{opt}}$ should be the one that satisfies $\phi\left(\lambda_{k,\mathrm{opt}}\right)=0$, which can be searched by the bisection method. Once $\lambda_{k,\mathrm{opt}}$ is obtained, the optimal solution to \eqref{GeneralADMMStep1} is then given by $\pmb{x}_{k,\mathrm{opt}} = \pmb{x}_k\left(\lambda_{k,\mathrm{opt}}\right)$ and $\tau_{k,\mathrm{opt}}=t_k\left(\lambda_{k,\mathrm{opt}}\right)$.

\end{document}